\documentclass[sigconf, anonymous=false, nonacm]{acmart}

\pdfoutput=1

\usepackage{pdfprivacy}
\usepackage{hyperref}
  \hypersetup{
    bookmarksnumbered,
    unicode,
    pdfencoding=auto,
    psdextra,
    breaklinks,
    linktoc=all,
    pdfcreator={},
    pdfproducer={},
  }

\usepackage{amsmath,amsfonts}%
\usepackage{array}
\newcolumntype{H}{>{\setbox0=\hbox\bgroup}c<{\egroup}@{}}
\usepackage{balance}

  \usepackage{bm}
\usepackage{booktabs}
\usepackage[noabbrev, capitalize]{cleveref} %
\usepackage{common}
  \tikzsetexternalprefix{./tmp/}
\usepackage{eurosym}
\usepackage{graphicx}
\usepackage{microtype}
\usepackage{multirow}
\usepackage{pdfcomment}
\usepackage{pifont}
\usepackage{relsize}
\usepackage{standalone}
\usepackage{stfloats}
\usepackage[caption=false]{subfig}
\usepackage{tabularx}
\usepackage[percent]{overpic}
\usepackage{wasysym}
\usepackage{xfrac}
\usepackage{xspace}

\newcommand{\her}{his/her\xspace}

\newcommand{\eg}{e.g.,\xspace} %
\newcommand{\cf}{cf.\xspace}   %

\newcommand{\code}[1]{\textsmaller[1]{\texttt{#1}}\xspace}

\newcommand{\xmark}{\ding{55}}%

\newcommand{\skipcite}[1]{\unskip\unpenalty}
\newcommand{\skipcitep}[1]{\skipcite{}}
\newcommand{\skipcitet}[1]{\skipcite{}}

\newcommand{\Euro}{\text{\euro}}

\newcommand{\tablesize}{}
\newcommand{\tablepad}{}%
\newcommand{\tablecappad}{\vspace*{-2mm}}
\newcommand{\figurecappad}{}%

\newcommand{\ourshorttitle}{Establishing Fast, Bidirectional\\Communication into Air-Gapped Systems} %
\newcommand{\ourtitle}{\ourmethod: \ourshorttitle}

\newcommand{\chris}{Christian Wressnegger}
\newcommand{\niclas}{Niclas Kühnapfel}
\newcommand{\nopx}{Maximilian Noppel}
\newcommand{\stefan}{Stefan Preu\ss ler}
\newcommand{\thomas}{Thomas Schneider}
\newcommand{\knrd}{Konrad Rieck}

\newcommand{\tub}{Technische Universität Berlin\xspace}
\newcommand{\ber}{\city{Berlin}\country{Germany}}

\newcommand{\tubs}{TU Braunschweig\xspace}
\newcommand{\bs}{\city{Braunschweig}\country{Germany}}

\newcommand{\kit}{Karlsruhe Institute of Technology}
\newcommand{\ka}{\city{Karlsruhe}\country{Germany}}

\newcommand{\purl}{https://intellisec.de/research/lasers}

\newcommand{\ourresult}[1]{{\vskip 10pt \noindent\bfseries #1.}}

\let\autoref\Cref

\newcommand{\todo}[1]{} %
\newcommand{\note}[1]{} %
\newcommand{\fix}[1]{} %

\newcommand{\yealink}{Yealink SIP-T21P E2\xspace}
\newcommand{\Yealink}{Yealink telephone\xspace}
\newcommand{\tlwr}{\mbox{TP-Link} \mbox{TL-WR1043ND}\xspace}
\newcommand{\tlmr}{\mbox{TP-Link} \mbox{TL-MR3020}\xspace}
\newcommand{\rpi}{Raspberry~Pi\xspace}

\begin{document}
\makeatletter
\if@ACM@anonymous
  \def\submissionID{334}
  \def\@acmSubmissionID{%
    \submissionID\vspace*{55pt}\gdef\@acmSubmissionID{\submissionID}}

  \thispagestyle{plain}
  \pagestyle{plain}
\fi

\if@ACM@nonacm
\newcommand{\mypadding}{10pt}
\newcommand\disclaimer{%
  \begingroup%
  \renewcommand\thefootnote{}\footnote{

\noindent This is a pre-print distributed via arXiv, the free
distribution service and open-access archive, as part of the Computing
Research Repository~(CoRR). It is posted here for your personal use not
for redistribution.\\\\
Submitted on June 8th, 2021 and revised on June 10th, 2021
\\\\
\textcopyright 2021, \url{https://intellisec.de}

  }%
  \addtocounter{footnote}{-1}%
  \endgroup%
}
\else
\newcommand{\mypadding}{0pt}
\newcommand{\disclaimer}{}
\fi
\makeatother

\title[\removelinebreaks{\ourtitle}]{\ourtitle\vspace{\mypadding}}
\pdfsuppressptexinfo-1
\pdfstringdefDisableCommands{%
  \def\vspace#1{}%
}

\author{\niclas}
\affiliation{%
  \institution{\tub}
  \ber
}

\author{\stefan}
\affiliation{%
  \institution{\tubs}
  \bs
}

\author{\nopx}
\affiliation{%
  \institution{\kit}
  \ka
}

\author{\thomas}
\affiliation{%
  \institution{\tubs}
  \bs
}

\author{\knrd}
\affiliation{%
  \institution{\tubs}
  \bs
}

\author{\chris}
\affiliation{%
  \institution{\kit}
  \ka\vspace*{\mypadding}
}

\begin{abstract}

  Physical isolation, so called air-gapping, is an effective method for
  protecting security-critical computers and networks. While it might be
  possible to introduce malicious code through the supply chain, insider
  attacks, or social engineering, communicating with the outside world
  is prevented.
  Different approaches to breach   this essential line of defense have
  been developed based on electromagnetic, acoustic, and optical
  communication channels.
  However, all of these approaches are limited in either data rate or
  distance, and frequently offer only exfiltration of data.
  We present a novel approach to infiltrate data to and exfiltrate data
  from \mbox{air-gapped} systems without any additional hardware
  on-site. By aiming lasers at already built-in LEDs and recording their
  response,
  we are the first to enable a long-distance (25\,m),
  \emph{bidirectional}, and fast (18.2\,kbps~in~\&~100\,kbps~out) covert
  communication channel. The approach can be used against any office
  device that operates LEDs at the CPU's GPIO interface.\disclaimer

\end{abstract}




\maketitle

\section{Introduction}

Individual devices, computers, or entire networks in high-security
environments are often physically isolated to prevent external access to
sensitive information. Such \emph{air-gapped systems} have neither wired
nor wireless network connectivity to the outside world and enforce
physical access control.
While this effectively prevents different types of network-based
attacks, in the past, we have seen several security incidents where such
systems have been successfully breached through attacks against the
supply chain~\citep{BasSaePil+19, Sha15}.
The importance of trustworthy software and hardware supply is underlined
by governmental restrictions on using foreign technology for critical
infrastructure in the US~\citep{web:uswh20} and
Europe~\citep{web:reuters20}.

Incidents at SolarWinds~\citep{web:fireeye20} and
CodeCov~\citep{web:codecov21} have recently shown the feasibility of
attacks against the \emph{software} supply-chain and the magnitude of
the consequences.
After a successful compromise of an air-gapped system, however, an
adversary faces the problem of interacting with the malicious code
injected in the device. As the isolation impedes regular communication,
alternative means are necessary for transmitting and receiving data.
Moreover, an attacker can only use hardware that is available on
\mbox{site---applying} additional equipment is not an option.
Academic research has explored different ways of attacking air-gapped
systems through optical~\citep{GurZadElo17, SugCyrRam+20},
acoustic~\citep{GurKacHas+15, GurSolDai+17a, GenPatSch+19},
thermal~\citep{GurMonMir+15, Gur19}, or even
electromagnetic~\citep{GurZadElo20} and
power-dependent~\citep{GurZadByk+20} communication channels.  Most of
these approaches only enable a unidirectional communication, that is,
either data infiltration or exfiltration. An authentic attack, however,
requires bidirectional communication to establish a command and control
channel, update the malicious functionality, or retrieve sensitive
information.
Despite the breath of prior work, existing covert channels hardly
address these practical constraints and fail to provide bidirectional,
efficient communication capabilities.

In this paper, we present \ourmethod, a novel attack vector that allows
to breach air-gapped boundaries \emph{and} \mbox{overcome} large
distances at high data rates. Our covert channel leverages built-in LEDs
of office devices to establish a \emph{bidirectional communication},
easily bridging distances between buildings in industrial parks or
embassy districts.
While LEDs are designed to emit light and can thus unnoticeably encode
information through high-frequency flickering, their ability to also
perceive light is largely unknown in the security community. In
particular, by directing a laser on the LEDs of office devices, we
induce a measurable current in the hardware that can be picked up by its
firmware and used to receive incoming data.
In contrast to conventional visible light communication~(VLC),
establishing this bidirectional communication is technically more
challenging: We cannot deploy any additional receiving equipment at the
device and need to operate the channel entirely from remote.
This bidirectional channel is applicable to devices where existing
LEDs are connected to a general purpose~I/O (GPIO) interface and hence
information can be sent and received through the device's firmware.

To demonstrate the efficacy of our attack, we systematically evaluate
the sending and receiving capabilities of LEDs and characterize the
needed laser modules with respect to power capabilities, wavelength, and
modulation. We find that for most devices a blue/purple engraving laser
for merely \SI{150}{\Euro} is sufficient to conduct the attack against
LEDs of different color.
For the back-channel, we investigate the suitability of cameras as found
in modern smartphones as well as specialized avalanche photodetectors.
In our experiments, we demonstrate infiltration and exfiltration of data
over~\SI{25}{\meter} at effective data rates.
While exfiltrating data using flashing LEDs has been investigated
before, we are the first to enable fast, \emph{bidirectional} covert
communication in this setting.\\[-7pt]

\noindent
In summary, we make the following contributions:
\begin{itemize}\setlength{\itemsep}{3pt}
\item\textbf{Novel infiltration technique.} We present a novel method
  for sending data towards air-gapped devices by utilizing
  built-in LEDs as receivers. Note, {LEDs are designed to emit
  light, rather than receiving it}.
  By using high-intensity lasers, we induce signals across tens of
  meters. %

\item\textbf{Significantly faster communication.}  In comparison to
  related work on covert channels, we are increasing the data rate of
  communication by an order of magnitude. For exfiltration, we improve
  the data rate by a factor of \num{25} and realize a speed-up factor
  of \num{110} for data infiltration.

\item\textbf{Practical implementation.} We demonstrate the feasibility
  of our bidirectional communication channel in different practical
  scenarios. We are able to bridge {\SI{25}{\meter}} at
  {\SI{18.2}{\kbps}} for infiltrating and {\SI{100}{\kbps}} for
  exfiltrating~data using LEDs already build in office devices.
\end{itemize}

The rest of the paper is structured as follows: We start with a
discussion on related work in \autoref{sec:related}, before we present
our attack and the newly proposed covert channel in
\autoref{sec:overview}. In \autoref{sec:attack-send,sec:attack-receive},
we then demonstrate data infiltration and exfiltration in practice.
\autoref{sec:conclusions} concludes the paper.

\begin{table*}[t]
  \tablepad\vspace*{-1mm}
  \caption{Overview of different covert channels. The first
  column
    indicates whether exfiltration~(\exfi), infiltration~(\infi), or
    both~(\both) is supported. The last column indicates whether
    increased privileges are necessary (\no), beneficial (\meh), or not
    required (\yes).}
  \label{tab:overview}
  \tablecappad

  \begin{center}
  {\tablesize

\newcommand{\mymidruleA}{\cmidrule(rl){1-3}\cmidrule(rl){4-6}\cmidrule(rl){7-11}}
\newcommand{\mymidruleB}{\cmidrule(rl){2-3}\cmidrule(rl){4-6}\cmidrule(rl){7-11}}
\newcommand{\dblwidth}{0.8pt}
\newcommand{\mydblmidruleB}{\cmidrule[\dblwidth](rl){2-3}\cmidrule[\dblwidth](rl){4-6}\cmidrule[\dblwidth](rl){7-11}}
\newcommand{\los}{{``line of sight''}}
\newcommand{\negl}{{``negligible''}}
\newcommand{\short}{{``short''}}

\newcommand{\maxv}{} %
\newcommand{\pfff}{{\hspace*{11pt}$<$~\tabbps{1}}}

\newcolumntype{C}[1]{>{\centering\arraybackslash}p{#1}}

\newcommand{\fnm}{\,\textsuperscript{a}}
\begin{tabular}{
    l
    c
    >{\hspace*{-2mm}}
    l
    lll
    S[table-format=3.1,table-space-text-post={\tabm{}}]
    S[table-format=4,table-space-text-post={\tabbps{}}]
    S[table-format=2.1,table-space-text-post={\perc{}}]
    l
    >{\hspace*{-1mm}}
    C{7mm}
  }
  \toprule
  &
  \multicolumn{2}{l}
  {\textbf{Method}}       &
  {\textbf{Channel}}      &
  {\textbf{Sender}}       &
  {\textbf{Receiver}}     &
  {\textbf{Distance}}     &
  {\textbf{Data rate}}    &
  {\textbf{BER}}          &
  {\textbf{Mode}}         &
  {\textbf{User}}         \\
  \mymidruleA
  \parbox[t]{2mm}{\multirow{5}{*}{\rotatebox[origin=c]{90}{other}}}
  & \exfi & PowerHammer~\citep{GurZadByk+20}            & power       & PC         & External probe  &{\negl\fnm}          & \tabbps{   200}   &  0\perc{}   & B-FSK & \yes \\%& \bad \\
  \mymidruleB
  & \exfi & ODINI~\citep{GurZadElo20}                   & magnetic    & PC         & External sensor & \tabm{1.5}          & \tabbps{    40}   & 10\perc{}   & OOK   & \yes \\%& \good \\
  \mymidruleB
  & \exfi & HOTSPOT~\citep{Gur19}                       & temperature & PC         & External sensor & \tabm{0.5}          & \pfff             & \na         & OOK   & \yes \\%& \good \\
  \mymidruleB
  & \both & BitWhisper~\citep{GurMonMir+15}             & temperature & PC         & PC              & \tabm{  9}          & \pfff             & \na         & TIS   & \yes \\%& \good \\
  \mymidruleA
  \parbox[t]{2mm}{\multirow{12}{*}{\rotatebox[origin=c]{90}{acoustic~~}}}
  & \exfi & \citet{Des14}                               & ultra-sound & Speaker    & Microphone      & \tabm{ 25}          & \tabbps{     8}   & 10\perc{}   & FSK   & \yes \\%& \good \\
  &       &                                             &             &            &                 & \tabm{  0}          & \tabbps{   345}   &  0\perc{}   & FSK   & \yes \\%& \good \\
  \mymidruleB
  & \both & \textsc{Mosquito}~\citep{GurSolElo18, GurSolElo20a}
                               \skipcite{GurSolDai+17b} & ultra-sound & Speaker    & Speaker         & \tabm{  3}          & \tabbps{   166}   &  1\perc{}   & B-FSK & \no  \\%& \good \\
  &       &                                             &             &            &                 & \tabm{  9}          & \tabbps{    10}   &  1\perc{}   & B-FSK & \no  \\%& \good \\
  \mymidruleB
  & \exfi & DiskFiltration~\citep{GurSolDai+17a}        & sound       & HDD        & Microphone      & \tabm{  2}          & \tabbps{     3}   &  0\perc{}   & OOK   & \no  \\%& \good \\
  \mymidruleB
  & \exfi & Fansmitter~\citep{GurSolElo20b}             & sound       & Fan        & Microphone      & \tabm{  8}          & \tabbps{     1}   & 10\perc{}   & B-FSK & \no  \\%& \meh  \\
  \mymidruleB
  & \exfi & AirHopper~\cite{GurKedKacElo14,GurMonElo17} & radio       & Std Video & FM receiver      & \tabm{  7}          & \tabbps{    80}   &  2\perc{}   & DTMF  & \yes \\%& \good \\
  &       &                                             & radio       & Ext Video & FM receiver      & \tabm{ 22}          & \tabbps{    80}   &  1\perc{}   & DTMF  & \yes \\%& \good \\
  \mymidruleB
  & \exfi & GSMem~\citep{GurKacHas+15}                  & radio       & RAM bus    & GSM Phone       & \tabm{1.1}          & \tabbps{     2}   &  6\perc{}   & B-ASK & \yes \\%& \good \\
  &       &                                             & radio       & RAM bus    & HW receiver     & \tabm{  3}          & \tabbps{  1000}   &  0.1\perc{} & FSK   & \yes \\%& \good \\
  &       &                                             & radio       & RAM bus    & HW receiver     & \tabm{ 30}          &{\na\footnote[1]{}}& \na         & \na   & \yes \\%& \good \\
  \mymidruleB
  & \exfi & USBee~\citep{GurMonElo16}                   & radio       & USB        & HW receiver     & \short              & \tabbps{   640}   & \na         & B-FSK & \yes \\%& \good \\
  \mymidruleA
  \parbox[t]{2mm}{\multirow{10}{*}{\rotatebox[origin=c]{90}{optical}}}
  & \both & aIR-Jumper~\citep{GurByk19}                 & light       & IR-LED     & camera          &{\los\fnm}           & \tabbps{    20}   & \na         & OOK   & \yes \\%& \good \\
  &       &                                             &             & IR-LED     & camera          &{\los\fnm}           & \tabbps{    40}   & \na         & ASK   & \yes \\%& \good \\
  \mymidruleB
  & \exfi & xLED~\citep{GurZadDai+18}                   & light       & LED        & camera          & \na                 & \tabbps{   120}   & \na         & OOK   & \no  \\%& \good \\
  &       &                                             &             & LED        & PD              & \na                 & \tabbps{  3555}\fnm& 5\perc{}   & OOK   & \no \\%& \good \\ %
  \mymidruleB
  & \exfi & LED-it-GO~\citep{GurZadElo17}               & light       & HDD LED    & camera          & \tabm{  8}          & \tabbps{   120}   & \na         & OOK   & \meh \\%& \good \\
  &       &                                             &             & HDD LED    & PD              & \tabm{  5}          & \tabbps{  4000}   & \na         & OOK   & \meh \\%& \good \\
  \mymidruleB         
  & \exfi & CTRL-ALT-LED~\citep{GurZadByk+19}           & light       & Keyboard   & camera          & \tabm{ 10}          & \tabbps{    30}   &  1\perc{}   & OOK   & \yes \\%& \meh  \\
  &       &                                             &             & Keyboard   & PD              & {\na}               & \tabbps{  2697}   &  8\perc{}   & OOK   & \yes \\%& \meh  \\
  \mydblmidruleB
  & \both & \textbf{\ourmethod}                         & light       & Laser      & LED             &\maxv\tabm{30}   &\maxv\tabbps{ 18200}&\maxv0\perc{}   & PWM   & \no  \\%& \good \\
  &       &                                             & light       & LED        & APD             &\maxv\tabm{25}   &\maxv\tabbps{100000}&\maxv0.1\perc{} & OOK   & \no  \\%& \good \\
  \bottomrule\\[-1pt]
  \multicolumn{5}{l}{Covert channel, but no data transmission:} \\[3pt]
  \toprule
  &
  \multicolumn{2}{l}
  {\textbf{Method}}       &
  {\textbf{Channel}}      &
  {\textbf{Sender}}       &
  {\textbf{Receiver}}     &
  {\textbf{Distance}}     &
  &
  &
  &
  {\textbf{User}}        \\
  \mymidruleA
  & \exfi & Synesthesia~\citep{GenPatSch+19}            & sound    & Screen     & camera             & {*}                 &                   &             &       & \yes \\%& \good \\
  & \infi & Light Commands~\citep{SugCyrRam+20}         & light    & Laser      & Microphone         & \tabm{110}          &                   &             &       & \yes \\%& \meh  \\
  \bottomrule\\[-3pt]
    \multicolumn{11}{l}
    {\fnm~~Determined theoretically or derived from unrepresentative settings (\eg power cord w/o branches, extraplotated from a few bits, \ldots).}
\end{tabular}

\ifstandalone
\bibliographystyle{abbrvnat}
\bibliography{../bib/other}
\fi
}
  \end{center}
  \tablepad\vspace*{-1mm}
\end{table*}

\section{Related Work}
\label{sec:related}

There exist various hardware-based attacks~\citep[see][]{NSA08}
facilitated by supply-chain compromise~\citep{BasSaePil+19}, which
inevitably leave physical evidence behind (the device itself). In this
paper, we thus focus on \emph{software} supply-chain
attacks~\citep[\eg][]{web:fireeye20, web:codecov21} that use covert
channels without any additionally brought-in equipment.
Our method enables bidirectional communication into air-gapped systems
using optical transmission and, thus, operates on the intersection of
two different fields of research: a)~Covert channels to exfiltrate and
infiltrate data, and b)~visual light communication, that forms the
foundation of our attack. Subsequently, we discuss both in detail.

\renewcommand{\paragraph}[1]{{\vskip 5pt \noindent\it #1.}}
\newcommand{\citeYear}[1]{\citeyear{#1}~\citep{#1}}

\subsection{Covert Channels of Air-Gapped Systems}

Academic research has investigated a variety of different approaches to
establish communication channels in and out of air-gapped systems.
\autoref{tab:overview} summarizes the most important ones.
Next to the specific transmission channel one may categorize covert
channels in 1)~methods that establish generic data transmission, and
2)~approaches that retrieve or send a very specific kind of information.

\subsubsection{Data Transmission}

There are multiple ways that are suitable for attacking air-gapped
systems. Subsequently, we describe the most prevalent ones that have
proven to be actionable in the~past:

\paragraph{Power consumption and magnetism} Recently, it has been shown
that it is possible to generate patterns in a system's power consumption
by controlling the workload of the CPU. These patterns, in turn, can be
measured on the wire with an external probe~\citep{GurZadByk+20}.
Unfortunately, the overall range is not well defined, but has been
measured on a straight power cord. Empirically, the authors have however
been able to reach a data rate of \SI{200}{\bps}.
Moreover, similar techniques may be used to produce magnetic fields to
encode data signals~\citep{GurZadElo20} that can be measured with
magnetic sensors. Thereby it is even possible to escape Faraday shields,
although the range of \SI{1.5}{\meter} is relatively short and the
achievable data rate with \SI{40}{bps} low.
Both techniques focus on unidirectional exfiltration of data.

\paragraph{Temperature} BitWhisper~\citep{GurMonMir+15} allows two
computers to bidirectionally transmit data to each other, encoded as
temperature differences caused by the system and measured with internal
sensors.
Surprisingly, this transmission could be sensed in a distance of
\SI{9}{\meter}, but merely with a data rate of less than \SI{1}{\bit}
per minute. This has later been used to show a unidirectional channel
with a customary smartphone as receiver~\citep{Gur19} over half a meter
at \SI{0.02}{\bps}. Here, transmission speed and distance has not been
the focus, but the fact that the attack is possible in a ``walk-by''
scenario.

\paragraph{Acoustic} As one of the first, \citet{Des14} demonstrates
the possibility of acoustic side-channels. In particular ultra-sonic
sound has been used to unidirectionally transfer data from ordinary loud
speakers to a microphone. More recently,
\textsc{Mosquito}~\citep{GurSolElo18, GurSolElo20a} has even turned
loudspeakers into microphones and enables bidirectional communication
this way. Unfortunately, the data rate is bound to \SI{10}{\bps} at
\SI{9}{\meter} only.
Moreover, it has been shown, that it is possible to generate sound in
the audible frequency spectrum using hard disks~\citep{GurSolDai+17a} as
well as fans~\citep{GurSolElo20b}. These, however, again are used for
exfiltration only at merely~\mbox{\num{1}--\SI{3}{\bps}}.

AirHopper~\citep{GurKedKacElo14, GurMonElo17} and
GSMem~\citep{GurKacHas+15} follow a similar route, but use radio signals
on different bands for the covert channel.
While the first demonstrates a video card's ability to send FM signals
for data transmission up to \SI{80}{\bps}, the latter uses memory
controllers to generate radio signals in the GSM band that can be
received using customary smartphones. While the method is argued to
bridge \SI{30}{\meter}, data rates could only be successfully measured
at a tenth of the distance at \SI{1000}{\bps}.
Subsequently, the same authors investigate how USB ports may be used to
produce RF signals in combination with dedicated hardware as
receiver~\citep{GurMonElo16}. Consequently, this and the methods above
rely on unidirectional communication.

\paragraph{Optical} More closely related to our findings, several
authors experiment with light-emitting diodes in various manifestations.
Modulation is achieved either directly by switching the LEDs of
routers~\citep{GurZadDai+18} or keyboards~\citep{GurZadByk+19} on and
off, or indirectly, for instance, by writing to the hard disk to make
the status LED flicker~\citep{GurZadElo17}.
With \SI{120}{\bps}, xLED~\citep{GurZadDai+18} is the fastest in this
setting. The specific value, however, has been derived from the
Nyquist–Shannon sampling theorem~\mbox{\citep[see][]{KenDav92}} and
resembles a theoretical upper limit for high-speed cameras at
\SI{240}{fps}. In \autoref{sec:exfiltrate-cam}, we verify this bound
empirically.
Additionally, the authors present data rates that may be yield with
photodetectors, that have been extrapolated from the maximum blinking
frequency the router has been able to~reach.

In contrast to the above, which only support data exfiltration,
\citet{GurByk19} demonstrate that security cameras that are equipped
with infrared LEDs can also be used to establish a bidirectional covert
channel. aIR-Jumper allows for \mbox{\num{20}--\SI{40}{\bps}} as long as
a direct line of sight is given. Measurements on the exact distance have
not been conducted.
Our approach extends this line of research by describing a novel
infiltration technique and demonstrating the practicability of
bidirectional communication at significantly higher data rates and
across large distances in a realistic environment.

\subsubsection{Inducing and Extracting Specific Signals}

A second line of research, addresses the exfiltration and infiltration
of application-specific signals.
For instance, \citet{GenPatSch+19} investigate the unintentional
emission of acoustic waves of electrical devices, such as computer
monitors. With Synesthesia, the authors present techniques to extract
screen contents by recording sound with built-in or external
\mbox{microphones---for instance} during a video call.
More recently, \citet{SugCyrRam+20} prove that photo-acoustic effects on
MEMS microphone diaphragms and photo-electric effects on the
microphone's ASIC may be used to induce sound via light. A modulated
laser light thereby appears as ordinary sound to the microphone.
This has been used to issue so-called ``light commands'' in speech
assistants. While extremely impressive, the method does not provide a
back-channel, meaning, it is possible to ask Alexa a question, but not
to hear her answer.
The attacker model and requirements for the attack, however, are similar to our work.

\begin{figure*}[htbp]
\centerline{
	\begin{overpic}[width=0.8\textwidth]{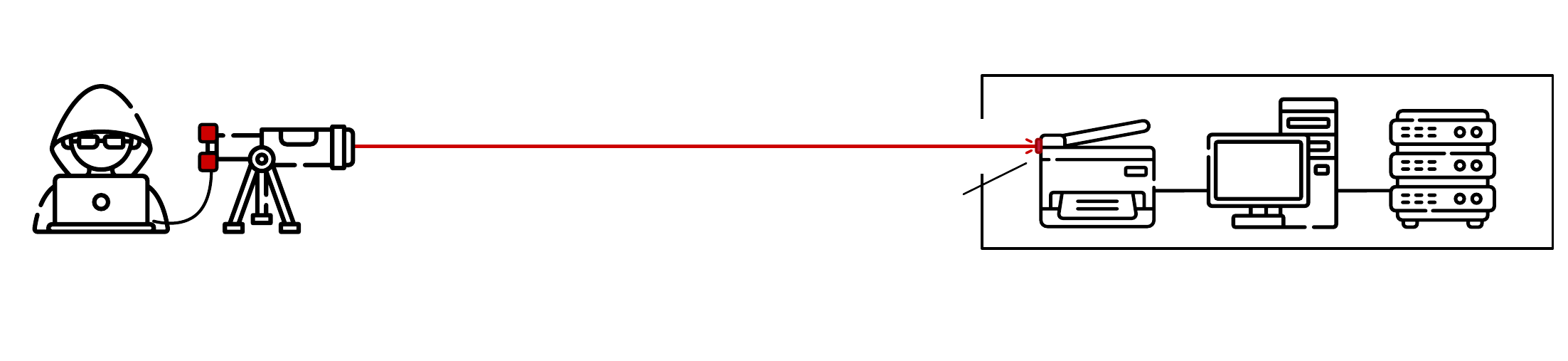}
      \put (31.0,15.25) {\footnotesize \colorbox{white}{\oneb Laser $\rightarrow$ LED (\SI{18}{\kbps})}}
      \put (27.0,11.00) {\footnotesize \colorbox{white}{\twob Photodetector $\leftarrow$ LED  (\SI{100}{\kbps})}}
      \put (1.5,4.5){\footnotesize\itshape \colorbox{white}{Adversary with laser and}}
      \put (5.5,2){\footnotesize\itshape \colorbox{white}{photodetector}}

      \put (35.10,3.25){\footnotesize \colorbox{white}{$>$\SI{20}{\meter} distance}}
      \put (57.20,8.90){\footnotesize \colorbox{white}{LED}}
      \put (72.00,19.7){\footnotesize \colorbox{white}{Air-gapped environment}}

      \put (64.7,4.5){\footnotesize\itshape \colorbox{white}{Compromised}}
      \put (65.3,2){\footnotesize\itshape \colorbox{white}{office device}}

      \put (85.0,4.5){\footnotesize\itshape \colorbox{white}{Isolated network}}
      \put (86.2,2){\footnotesize\itshape \colorbox{white}{infrastructure}}
    \end{overpic}
}
\figurecappad%
\caption{Schematic depiction of a bidirectional covert channel using
  an device's LED. \oneb~Inbound communication~(data infiltration) is
  established through a laser. \twob~Outbound communication~(data
  exfiltration) is received by an avalanche photodetector.}%
\label{fig:setup_schematic}
\end{figure*}
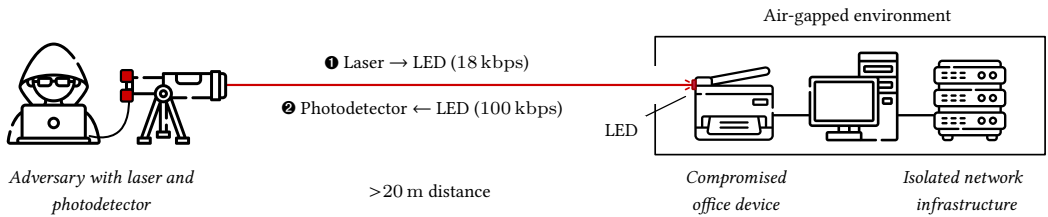

\subsection{Visible Light Communication (VLC)}

The communication with light is a well explored field of application in
photonics. Data is transmitted by modulating light in the visible
spectrum from \SI{380}{nm} to \SI{750}{nm}~\citep{Jovicic13, Pathak15}.
There exist several similar technologies, such as optical wireless
communication~(OWC)~\citep{Uysal14}, free-space optical
communication~(FSO)~\citep{Chan06}, and light
fidelity~(Li-Fi)~\citep{Haas16}. For a discussion on the similarities
and differences of these types of communication systems, we refer the
interested reader to the survey by~\citet{Matheus19}.

The transmitter in visible light communication (VLC) systems usually is
an light-emitting diode (LED), while two different types of receivers
may be used to capture the transmitted signal: First, a photodetector
also referred to as photodiode or non-imaging receiver, and second, a
camera or imaging sensor~\citep{Pathak15}. Both have been considered in
our evaluation of the presented covert channel.
Visible light communication comes with certain limitations, for
instance, the obvious need for a direct line of sight and the fact that
the achievable data rate falls abruptly with increasing
distance~\citep{Kahn97}.
Nevertheless, state-of-the-art systems with a single dedicated LED and
on-off-keying can reach transmission speeds of up to \SI{10}{\Gbps} in a
distance of~\SI{1.6}{\meter}~\citep{KomNak04}.

For this, however, VLC systems make use of elaborate optical equipment
at the receiving end (in our case the targeted device) to bundle,
stabilize, and focus the communication signal. This of course is not
possible for the attack scenario considered in this paper, as we breach
\emph{unmodified, air-gapped consumer devices} using their already
built-in LEDs.
Consequently, we face a more difficult application that is off the
usually studied techniques in visual light communication.

\iftrue %

\section{Bridging the Air-Gap}
\label{sec:overview}

As the name \emph{light-emitting diode} suggests, LEDs are designed and
built to send out light. In office devices they are primarily used to
indicate a device's state or realizing small displays. The fact that
LEDs may also be used as a receiver~\citep{SteKowMak+15}, however, is
widely unknown.
We make use of this property to establish a bidirectional communication
channel between an attacker and the compromised software of an
air-gapped device using its built-in LEDs.

\autoref{fig:setup_schematic} depicts the main principle of our attack:
\oneb~Using a strongly focused laser beam, current is induced in the LED
of a device. If the device operates the LED at a general-purpose I/O
interface, the corresponding voltage can be measured by the firmware
and used to transmit and thus infiltrate data. \twob~To exfiltrate data
the device flashes the LED, such that the attacker is able to observe
the light with a telescope similar to the one used to focus the laser.
While a direct line of sight is necessary, glass windows do not obstruct
the light to an extent that would counteract the attack.
Also, it is important to note that the device's LEDs are still
functioning properly for their primary purpose (\eg signaling device
state) before and after transmission.

In \autoref{sec:attacker}, we proceed to specify the considered attacker
model, before we outline the necessary equipment of the adversary in
\autoref{sec:infiltrate,sec:exfiltrate}, for infiltrating and
exfiltrating data, respectively. In \autoref{sec:targets}, we detail the
class of devices that are attackable and derive a suitable communication
protocol in \autoref{sec:protocol}.

\paragraph{\normalfont\bfseries Example} To assist the
subsequent description of the attack, we use the \yealink telephone
as a reoccurring example throughout the paper.
For visually indicating the device's state, this telephone makes use of
two paired green and red SMD\footnote{\label{fnote:smd}Surface-mounted
device~(SMD)} LEDs, and another individual SMD LED that emits red light.
We refer to the paired diodes as \code{green} and \code{red-1}, and the
individual one as \code{red-2}.
These LEDs do not only differ in the emitted light's wavelength, but
also to which wavelength they react, when hit by a high-intensity light
beam. As an example, \autoref{fig:ranges}~(top) shows the sensitivity of
the first red and the green diodes of the \Yealink to incoming light of
different wavelength. A thorough characterization of a wide range of
LEDs is provided in \autoref{sec:leds-send,sec:leds-recv} when
discussing data infiltration and exfiltration in practice.

\subsection{Attacker Model}
\label{sec:attacker}

For our attacker model, we assume that an initial compromise has
happened on the target device through the software supply-chain similar
to the incidents at SolarWinds~\citep{web:fireeye20} and
CodeCov~\citep{web:codecov21}. For example, a regular update of the
device's firmware might unnoticeably add the necessary code for sending
and receiving data through a built-in LED.
While many office devices, such as desk telephones and printers, expose
a vast attack surface to the outside world, we assume adequate
isolation, that in further consequence, poses the necessity for bridging
the air gap.
Moreover, we assume that device characteristics, such as built-in LEDs
and circuit details, are known to the attacker. This, however, is
information that can be derived from identically constructed devices and
device drivers.
Finally, for the attack to succeed, a direct line of sight is necessary
to observe and actuate the LED. Note, that for reading and writing the
general-purpose IO (GPIO) interface usually system privileges are
required.
{The adversary, however, neither needs physical access to the
device, nor requires the owner of the hardware to accidentally or
intentionally interact with the device. Moreover, we do not assume any
upfront hardware modifications of the device.}\\

\begin{mdframed}[style=HighlightFrame]
  \textbf{Remark.} While the transmission of data using the \ourmethod
  attack only takes a split second, the operation of high-intensity
  laser modules raises safety issues. We assume that the adversary a)
  operates the laser with great care to avoid endangering people in the
  vicinity and b) is willing to accept the remaining risk.
  \emph{We, however, conducted all experiments under strict safety precautions.} 
 \end{mdframed}%

\subsection{Infiltrating Data~\textnormal{(\oneb)}}
\label{sec:infiltrate}

In contrast to the targeted device, the adversary is entirely free to
choose the hardware necessary to establish a stable communication on the
attacker's end of the covert channel. While there are virtually no upper
limits with respect to expense and size of the equipment, we implement
the attack with mobile consumer components that are easily available and
use household power. Our setup is shown in \autoref{fig:setup}.

The wavelength of the laser beam is crucial for the success of the
attack and needs to match the absorption range of the targeted LED. In
the bottom part of \autoref{fig:ranges}, we show the wavelength of four
different lasers that emit blue/purple (two spikes to the left), green
(middle spike) and red light (right most spike). This clearly shows that
not all lasers can be used for inducing current into any diode.
While the blue and purple lasers work well for the green, but not for
the red LED of the Yealink~telephone, for the green laser it is %
the other way around.
Interestingly, the red laser does merely scratch the range of the
red diode and is also far off the green~one.

The four laser modules vary greatly in power but all are
commercially available without restrictions in a range from
\num{5}--\SI{150}{\Euro}. Regular laser modules with less than
\SI{5}{\mW} are considered \mbox{class~\num{1}--\num{3}R} and are available
across rather broad light spectra in the range of
\num{405}--\SI{980}{\nm}. 
More powerful devices emit up to \SI{100}{\mW} and fall into
class~\num{3}B. In common usage, such lasers are only available for
fixed wavelengths, such as \SI{405}{\nm} (purple), \SI{532}{\nm}
(green) and \SI{650}{\nm} (red).
Engraving lasers, finally, reach multiple Watts of energy and fall
into class~\num{4}. These lasers typically use a wavelength of
\textasciitilde\SI{450}{\nm}.

The distance that can be bridged using such lasers depends on the
ability to focus the laser beam and the resulting optical power raised
at the LED. The better the used telescope, the larger the distance.
As part of our evaluation on data infiltration in practice, we provide
measurements for the exact optical power of different lasers considered
in our experiments and describe the used optics that are necessary in
\autoref{sec:lasers}.

\tikzsetnextfilename{led_absorption_sample}
\begin{figure}[t]
\centerline{
  \begin{tikzpicture}[
    font=\sffamily\scriptsize, every node/.style={fill=white, inner
    sep=0}
  ]
    \node at (0, 0)
    {\includegraphics[width=0.5\textwidth]{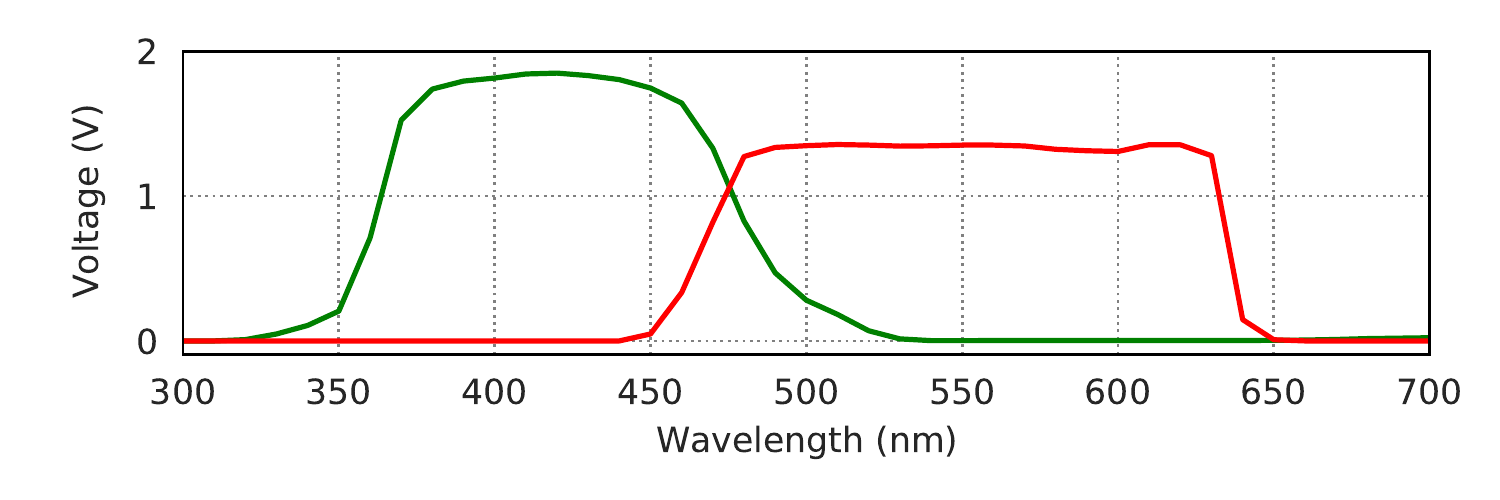}};
    \node[text=red]
      at ( 3.30, 0.10) {red-1};
    \node[text=green!50!black]
      at (-2.65, 0.75) {green};
  \end{tikzpicture}
}\vspace*{-3mm}
\centerline{\includegraphics[width=0.5\textwidth]{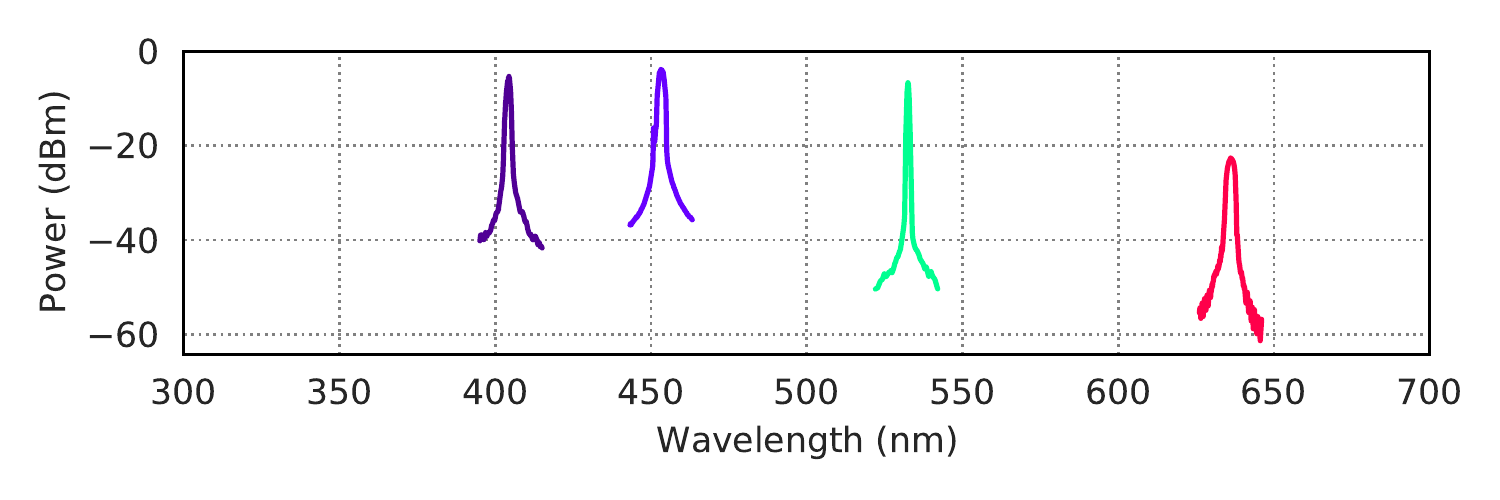}}
\figurecappad\vspace*{-2mm}
\caption{Absorption spectra of the \yealink's LEDs at
  \SI{25}{\milli\watt}~(top) and the emission spectra of two blue/purple, one
  green, and one red laser module~(bottom).}\vspace*{-2mm}
\label{fig:ranges}
\end{figure}

\begin{figure}[b]\vspace*{-4mm}
  \subfloat[Infiltration]
  {\includegraphics[width=0.48\columnwidth]{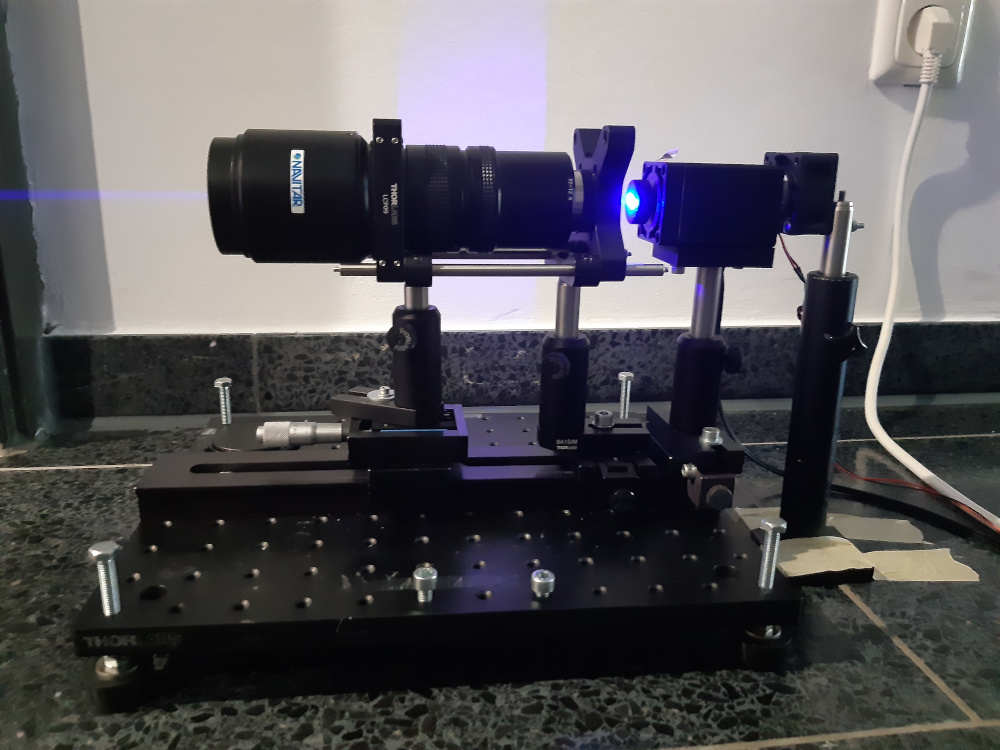}}
  \hfill \subfloat[Exfiltration]
  {\includegraphics[width=0.48\columnwidth]{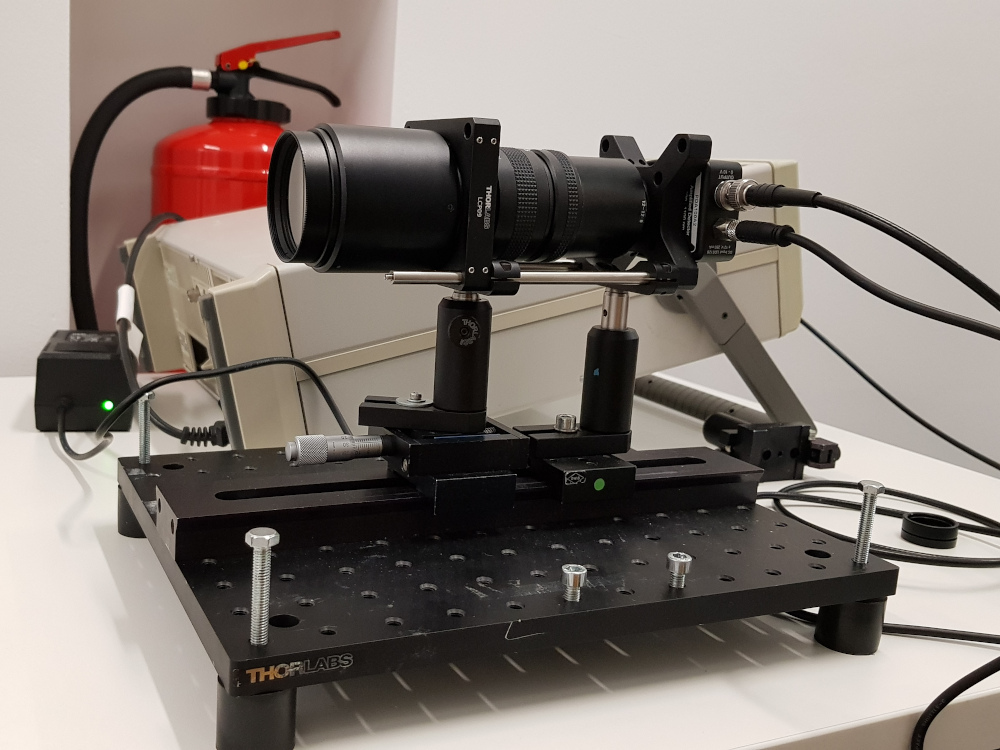}}
  \hfill
  \figurecappad%
  \caption{Telescopes, laser module, and oscilloscope used for a)~data
  infiltration and b)~data exfiltration.}
  \label{fig:setup}
\end{figure}

\subsection{Exfiltrating Data~\textnormal{(\twob)}}
\label{sec:exfiltrate}

In contrast to producing light beams with high precision, the
requirements at the attacker's end for recording light signals sent out
by the target device are less specialized.
In the most simple case, a modern smartphone with a consumer high-speed
camera, such as the iPhone since version~6~\citep{web:apple_iphone6}, is
sufficient. These cameras capture light at \SI{240}{fps} and thus enable
a moderate transmission rate. Furthermore, the bridgeable distance is
constraint by the sensitivity of the camera.
Both aspects can be improved upon by using specialized optics, such as a
telescope similar to the one used to focus the laser beam, and a
dedicated light sensor, such as a photodetector~(PD).
In~\autoref{sec:exfiltrate-cam,sec:exfiltrate-apd}, we evaluate both
scenarios in practice.

Conventional photodetectors have a response time of a few nanoseconds
only and a broad spectral response.
However, they are limited in sensitivity and have a small active area,
which makes their use for capturing light signals over large distances
difficult~\citep[see][]{GurZadElo17, GurZadByk+19}.
For measuring very small amounts of light, we thus make use of so-called
\emph{avalanche photodetectors~(APD)}. These detectors create a strong
electric field to increase the sensitivity to incoming light. When a
photon hits the sensor, this electric field accelerates the electrons
leading to the production of secondary electrons through impact
ionization. The resulting avalanche of electrons produces a gain factor
in the hundreds.
This amplification limits the usable bandwidth of the detector to
\SI{100}{\kHz}. Still, this rate significantly surpasses high-speed
cameras and enables us to outperform existing covert channels. As
part of our evaluation on data exfiltration in practice, in
\autoref{sec:photodiodes}, we characterize the sensitivity of
APDs in more detail.

\begin{figure*}[t]
  \vspace*{-4mm}
  \centerline{
  \subfloat[\hspace{-30mm}\label{fig:mr3020_schematic}]
  {\includestandalone[mode=buildmissing, height=30mm]{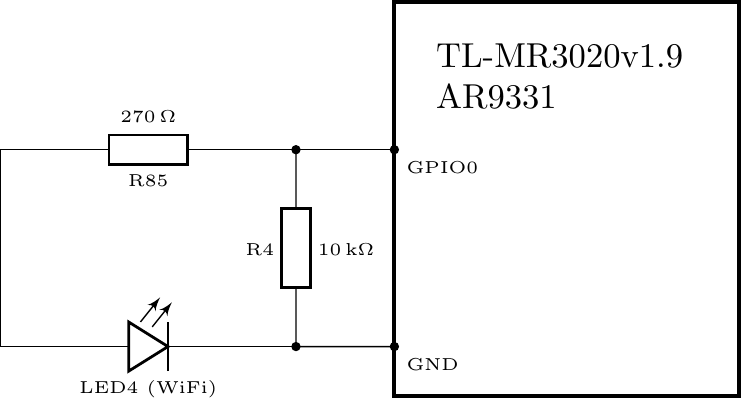}}
  \hfill
  \subfloat[\hspace{-22mm}\label{fig:wr1043nd_schematic}]
  {\includestandalone[mode=buildmissing, height=30mm]{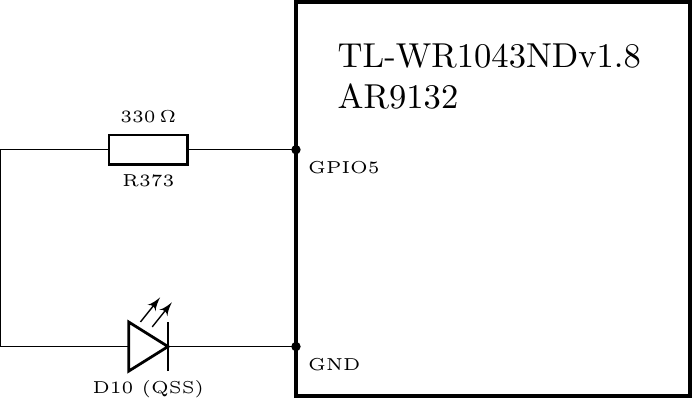}}
  \hfill
  \subfloat[\hspace{-30mm}\label{fig:t21p_schematic}]
  {\includestandalone[mode=buildmissing, height=30mm]{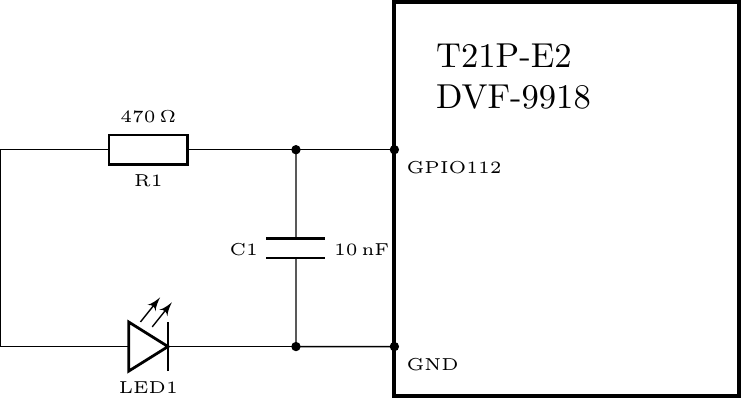}}
  }
  \figurecappad%
  \caption{Schematics of three attackable devices: a)~\tlmr mobile router,
  b)~\tlwr office router, c)~\mbox{\yealink} telephone. Each device
  uses a different circuit for connecting the LED to the GPIO interface.}
  \label{fig:schematics}
\end{figure*}

\subsection{Attackable Devices} %
\label{sec:targets}

A few prerequisites need to hold on a hardware level, such that a device
can be targeted using our attack. For detecting laser pulses an already
built-in LED needs to be connected to and directly driven by the
processor via a GPIO interface. Whether the LED is configured as
\emph{active high} or \emph{active low} is not relevant, as long as it
is \textbf{not}~permanently carrying electric current (constantly
glowing).
Also, while the presence of pull-up/down resistors or any additionally
used series-connected resistors affect the GPIO levels, they do not
impede the attack.
To determined the prevalence of devices with built-in LEDs usable for
the \ourmethod~attack, we have analyzed the device tree specifications
of the Linux kernel. \perc{48} (\num{679}) of \num{1394} investigated
boards use LEDs at the GPIO interface, from which the majority
(\num{522}) is operated in an active high configuration.

Some devices additionally use a capacitor between GPIO and ground, in
order to filter the signal. This does not impede the communications
channel either, but requires alternative sampling strategies at the
target device when receiving data to be infiltrated.
In the following experiments, presented in
\autoref{sec:attack-send,sec:attack-receive}, we thus consider three
representative office devices:%
~a)~\tlmr mobile router, b)~\tlwr office router, and c)~\yealink
telephone. Each of these devices uses a different circuit for connecting
their built-in LEDs to the general-purpose I/O interface of the CPU.
Further details on these devices are listed
in~\autoref{tab:pre_exp_rates}.

\autoref{fig:schematics} shows simplified schematics for all three.
For both routers the processor and LEDs are directly linked via a
series-connected resistor, while an additional capacitor is used in
Yealink's device to avoid interference. The additional resistor \texttt{R4} of the \tlmr needs to be present in case there is no internal pull-down. The CPU of the \tlwr, in turn, does implement a
pull-down resistor internally and, thus, can be used directly.
The design of the different circuits requires the use of two different
sampling~strategies that we detail in the following.

\subsubsection{Immediate sampling} In the case of a series-connected
resistor between LED and the processor's GPIO, the firmware's GPIO API
can be directly used to get the current state of the pin. Toggling the
state by pointing the laser at the LED requires injecting a sufficient
amount of energy to the circuit, such that the voltage at the
pull-down resistor reaches the minimal switching threshold.
For instance, for the \tlmr an induced current of
\SI{200}{\micro\ampere} is needed to reach a voltage of \SI{2}{\volt}
due to the built-in \SI{10}{\kilo\ohm} resistor.

\subsubsection{Delayed sampling} Immediate sampling is not possible if
there is an additional capacitor between GPIO and ground (irrespective
of being pulled up or down), because the injected energy is first stored
in the capacitor without toggling the GPIO.
Therefore, it is necessary to charge the capacitor until its voltage is
high enough to retrieve a sample by reading the state of the GPIO as
shown in \autoref{fig:ex_trans}.
The sampling method consists of three phases: First, the capacitor is
charged up to the desired voltage. Second, the logical GPIO value is
retrieved. After that it is necessary to unload the capacitor in the
third phase.
This process takes a constant amount of time and, thus, leads to a lower
sampling rate than the theoretically possible bandwidth of the LED.

Technically, the time $t$ required to charge the unit is defined by
its electrical capacitance $C$ and is computed as follows:
$t_c = \frac{C}{I} \cdot U_c$\,, where $I$ represents the induced
current and $U_c$ the voltage level at the capacitor in the target
device. An exemplary capacitor with \SI{10}{\nano\farad}, an applied
voltage of \SI{2}{\volt}, and an laser-induced current of
\SI{20}{\micro\ampere} thus requires $t = \SI{1}{\milli\second}$ to
charge, meaning, the communication protocol is delayed by that
duration.
The advantage of such delayed sampling, however, is that less energy
has to be injected to the LED, as it is possible to wait until the
capacitor is charged.
For the same reason, transmission is more robust in this setting as
charging may be subject to discontinuity as caused by vibrations
(\cf~\autoref{sec:lasers}) without impeding the attack.

\begin{figure}[b]
    \centering
    \includestandalone[mode=buildmissing,width=1.0\columnwidth]{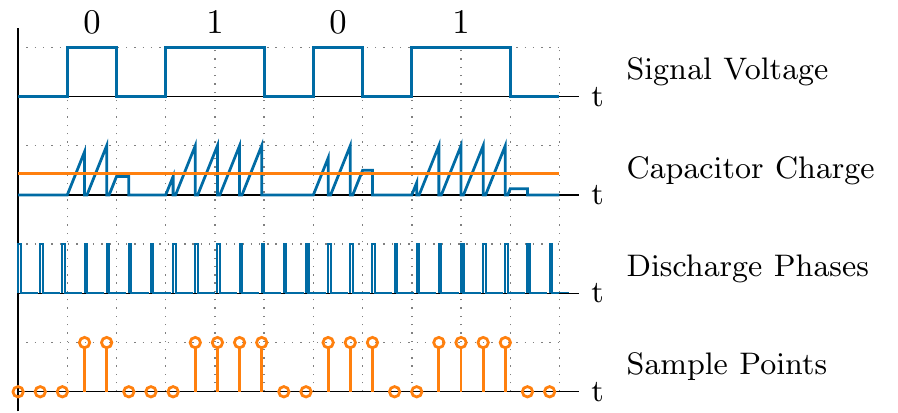}
    \vspace*{-5mm}
    \figurecappad
    \caption{Exemplary transmission in presence of \mbox{a capacitor}.}
    \label{fig:ex_trans}
    \label{fig:modulation}
\end{figure}

\subsection{Communication Protocol}
\label{sec:protocol}

Infiltrating data into devices without real-time capable processors
requires the use of an easy and robust modulation technique.
For our attack, we hence use a variant of pulse-width
modulation~(PWM)~\citep[see][]{KenDav92}:
Transmitting a zero bit corresponds to a short pulse while transmitting
a one bit is achieved by a long pulse as indicated
in~\autoref{fig:modulation}~(top). This scheme is owed to the sampling
strategy described above. A~high value indicates that the laser is
active, while a low value indicates that it is off. The slots where the
laser is not active are used to tell individual bits apart.
The achievable data rate consequently depends on the ratio of zero and
one bits. For subsequent experiments, we consider the worst-case (only
1-pulses are sent) to report a lower bound of the data~rate.

For exfiltrating data, however, we are not restricted to a particular
sampling strategy as the attacker may choose \her hardware at will at
the receiving end.
On the office device, we thus use classical on-off-keying (OOK)
for sending data, that is, a high value encodes a one bit, while a low
value represents a zero bit. The duration of transmitting each is
identical $\tone = \tzero$.
Separating bits as described above is not necessary: $\toff = 0$.
To further increase the data rate other encodings, such as
amplitude-shift keying~(ASK)~\mbox{\citep[see][Chp.~3]{KenDav92}} or
(binary) frequency-shift keying~(FSK,
B-FSK)~\citep[see][Chp.~5]{KenDav92} are possible. Exploring this,
however, is left to future work.

\section{Infiltrating Data}%
\label{sec:attack-send}

After outlining the attack setting and describing the underlying channel
for covert communication, we now demonstrate the attack in practice and
begin with infiltrating data into remote devices.
By directing a high-intensity laser beam onto a office device's LED it
is possible to induce a measurable current that allows to establish data
communication.
The experimental setup is detailed in \autoref{fig:setup_attack}.

\begin{figure}[htbp]
\centerline{\includegraphics[width=\columnwidth]{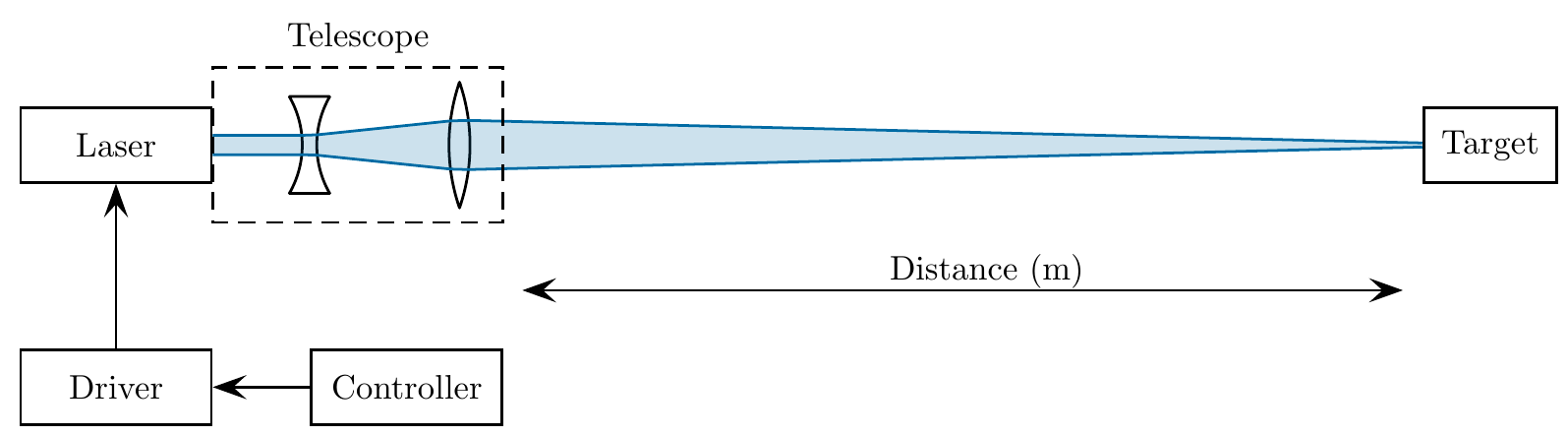}}
\figurecappad\vspace*{2mm}
\caption{Experimental setup for infiltrating data.}
\label{fig:setup_attack}
\end{figure}

In particular, in \autoref{sec:leds-recv}, we systematically evaluate
the light absorption characteristics of different LEDs that can be used
as a receiver and thus contrary to its intended purpose. In
\autoref{sec:lasers}, we then describe the laser modules necessary to
actuate the LEDs and address peculiarities of the used hardware, the
necessary optics, and issues with~vibrations. Based on these
characterizations, we conduct two experiments in
\autoref{sec:sending-results}. We first establish an empirical upper
limit for infiltrating data based on the described target devices over a
rather short distance, before we conduct measurements in an realistic
setting with distances of up to \SI{40}{\meter}.

\subsection{LEDs as Receiver}
\label{sec:leds-recv}

For the attack to succeed, the wavelength of the used laser needs to
align with the absorption spectrum of the LEDs to establish a reliable
communication channel for infiltrating data.
We thus inspect the light absorption of common LEDs, that enables us to
put the measurements of the device-specific LEDs as presented in
\autoref{fig:ranges} for the \Yealink into perspective.
Details and specifications of the specific LEDs are provided in
\autoref{app:leds}.

In principle, any reversely biased LED can act as a poorly designed
photodetector for which the flow of electricity is caused in the entire
active layer and the junction itself. Therefore, the absorption
spectrum of an LED equals the spectrum of the emitted wavelength.
Hence, we begin with determining the absorption spectra of the diodes
in question. This is commonly done using a white-light source (\eg~a
LOT Quantum Design, LSH~302) and a monochromator (\eg~MC~Pherson~2035)
to continuously adjust the wavelength. To better focus the light on
the LED, additional optics is used.
\autoref{fig:LED_Laser_Spectra} shows the absorption curves for seven
diodes, for which we measure the induced voltage in dependence on the
illuminated light for different wavelengths with a resolution of
\SI{5}{\nm}.

\tikzsetnextfilename{led_absorbtion_curves}
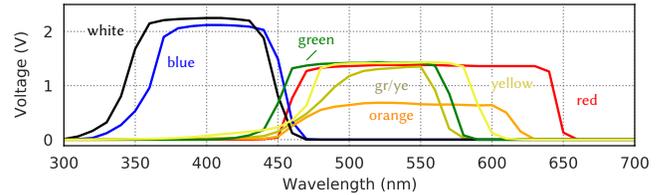
\begin{figure}[htbp]%
\centerline{
  \begin{tikzpicture}[
    font=\sffamily\scriptsize,
    every node/.style={fill=white, inner sep=0}
  ]
    \node at (0, 0) {
      \includegraphics[width=0.5\textwidth]{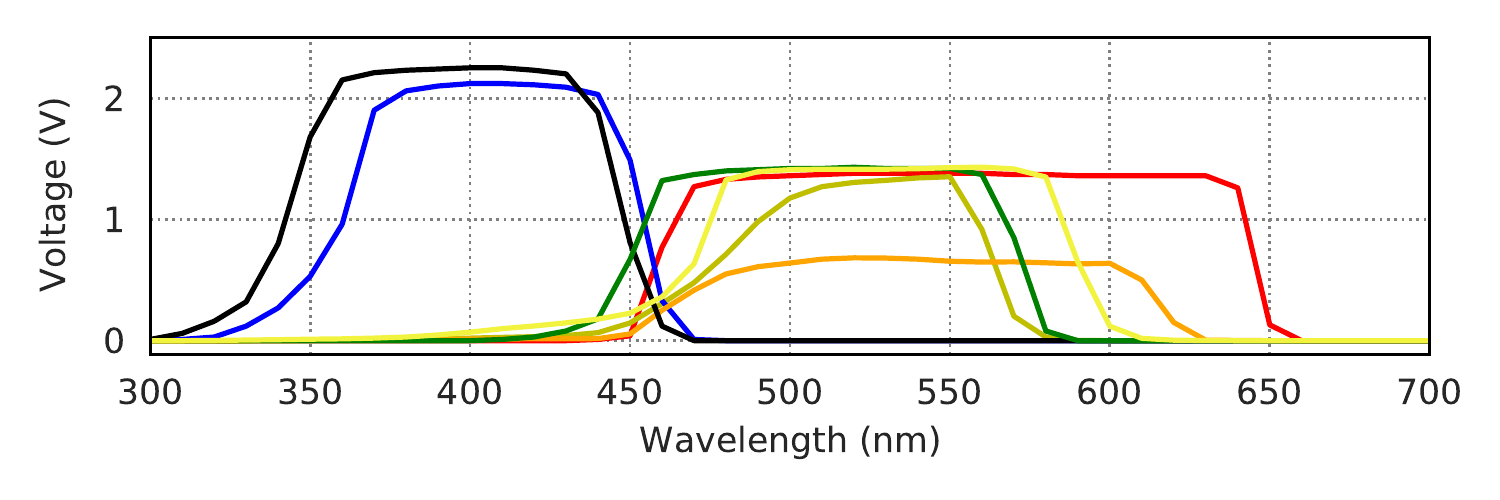}
    };
    \node
      at (-3.0,  0.92) {white};
    \node[text=blue]
      at (-2.00,  0.50) {blue};
    \node[text=red]
      at ( 3.4,  0.00) {red};
    \node[text=yellow!80!black]
      at ( 2.4,  0.20) {yellow};
    \node[text=orange]
      at( 0.8, -0.25) {orange};

    \node[text=yellow!50!black]
      at ( 0.8, 0.17) {gr/ye};

    \node[text=green!50!black]
      at (-0.2,  0.75) {green};
    \draw[green!50!black] (-0.2,  0.65) -- (-0.33,  0.52);
  \end{tikzpicture}
}
\figurecappad\vspace*{-2mm}
\caption{Absorption curves of seven LEDs of different color.}
\label{fig:LED_Laser_Spectra}
\end{figure}

Usually, the absorption curve is shifted towards the lower end of the
emitted spectra of wavelengths, as the LED is not able to detect photons
of lower energy than its band~gap~\citep{SteKowMak+15}. This is in line
with our measurements, where the absorption of most diodes ranges
broadly around the original color.
Interestingly, this is not the case for the green SMD LED that the
\Yealink uses (\cf~\autoref{fig:ranges} top), where the
absorption spectrum does not fit the emitting color at all, but leans
towards white and blue color.
Here, apparently, a white LED with a green colored cover has been
built-in rather than a diode that actually emits green light.
Of course, the diode may still be used for establishing a covert
channel, but needs to be illuminated with a blue laser instead.

\subsection{Laser Modules}
\label{sec:lasers}

The wavelength of the used laser beam is crucial for establishing a
communication channel with a particular LED. To provide further insight
into these relations, we determine the emission wavelengths and laser
spectra of different laser modules in \autoref{sec:laser-spectra}.
Next to a matching (and powerful) laser, it is crucial for a successful
attack to precisely focus the laser beam onto the targeted LED. In
\autoref{sec:optics}, we present the optical equipment used in our
experiment and discuss how to handle vibrations to stabilize the laser
beam.

\subsubsection{Laser spectra}
\label{sec:laser-spectra}

We measure the light of four different lasers using the free space input
of an optical spectrum analyzer (Anritsu~MS9701C) within a span of
\SI{20}{\nm} and a resolution of \SI{0.04}{\nm}.
The corresponding wavelengths are shown in the bottom part of
\autoref{fig:ranges}. As an example, to actuate the LEDs the \Yealink
uses, the peak in optical power needs to reside in the absorption range
of the diodes (top part of the~figure).
For bridging large distances, in turn, the optical power of the laser is
crucial. In \autoref{fig:laser_power}, we characterize the green
class~3B laser pointer and the purple/blue class~4 engraving laser out
of the four laser modules mentioned above.
For both, we measure the optical output power in dependence of the
driving current. While the green laser pointer reaches an output of up
to \SI{60}{mW} for the maximum current, much higher power is reached
with the engraving laser. In order to avoid damage of the optical sensor
the measurement has been limited to \SI{100}{mW}. According to the
specification, the laser may achieve up to \SI{6}{W}.

\begin{figure}[htbp]
  \vspace*{-3mm}%
  \centerline{\includegraphics[width=0.48\textwidth]{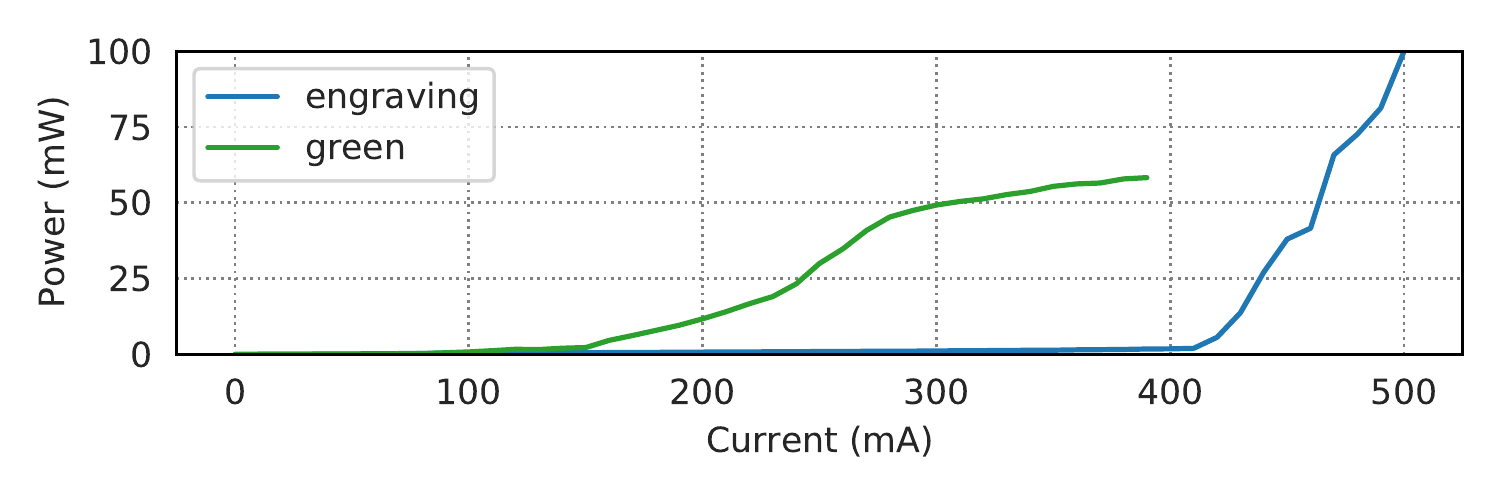}}
  \vspace*{-5mm}
  \figurecappad
  \caption{(Optical) laser output power over the input current.}
  \label{fig:laser_power}
\end{figure}

Moreover, to establish a communication protocol and transmit data, the
laser is directly modulated, that is, the optical power of the laser is
changed by varying the input current.
It is important to note, that direct modulation also changes the output
wavelength, which may cause signal loss at the edge of the absorption
curve.
Moreover, the modulation bandwidth of lasers crucially depends on the
speed of the driver controlling the current.
While integrated driver circuits for off-the-shelf laser pointers are
limited to a few \si{\kilo\hertz} in speed, much higher bandwidths can
be achieved with external current controllers.
In general, a directly modulated laser module can reach bandwidths of up
to~\SI{30}{\giga\hertz}~\citep[see][]{Tadokoro11}.%

\paragraph{\normalfont\bfseries Choice of laser module}
In order to transmit the highest possible optical power over a long
distance, subsequently, we make use of the strong class~\num{4}
engraving laser, which has been purchased for merely \SI{150}{\Euro} on
Ebay. To guard against any damage in the surroundings, we operate the
laser module in a strictly marked out area using protective gear.
The laser is driven by a separate current source (ILX Lightwave
LDC3744C) that provides stable output with redundant current limiters.
In this setting the laser current may reach up to \SI{4}{\ampere} and a
compliance voltage of~\SI{10}{\volt}.
We run the driver in constant current mode with external modulation such
that the laser is switched on and off during modulation. The modulation
signal itself is provided by an external controller.

\subsubsection{Stabilizing laser beams across distance}
\label{sec:optics}
\label{sec:vibrations}

The mentioned class~4 engraving laser comes with a small lens applied,
such that the light beam diverges after a few meters. To counteract
this, we use a telescope at the attacker's side to focus the beam in
distance.
We use the Navitar Zoom~7000 telescope, which is a close-focusing macro
video lens with a working distance from \SI{5}{inch} to (theoretically)
infinity and is parfocal over the entire zoom range.
However, in practice the beam of course shows some broadening over
particular long distances above \SI{40}{m}. With a price of
\SI{300}{\Euro} the used telescope is on the lower end of the
scale for professional optical equipment. Higher investments for better
optics likely enable to extend the practical parfocal range
significantly.

To overcome long distances it additionally is crucial to handle
vibrations and stabilize the laser beam. With increasing distances,
small vibrations lead to significant movement of the light spot. Minor
movements of the building, for instance due to outside traffic, people,
and elevators, may thus impede the attack, although unnoticeable as
a~person.
To stabilize the setup, we mount the laser and the telescope on an
optical plate with shock absorbing feet (Thorlabs~AV4) made out of
Sorbothane, a synthetic viscoelastic urethane polymer. For the given
load, the absorption efficiency is \mbox{\num{80}--\perc{100}} for
frequencies above \SI{49}{\hertz}, which is sufficient for our
experiments.
To further compensate for vibrations one may also use fast steering,
Piezo-activated
mirrors%
that counteract movements in (near) real-time. %
Another significant source of distortions are movable parts of the
sending equipment itself, such as fans. %
For our experiments, we hence detach the fan from the laser's heat~sink.

Once appropriate measures against vibrations have been established, the
focal point of the telescope is adjusted on the target---the device's
LED---for the respective distance and applied with ultra-fine adjustment
screws.

\begin{table*}[t]
  \tablepad
  \caption{Data rates for infiltrating data into the target device by inducing current into LEDs using high-intensity laser beams.}
  \label{tab:recv}
  \tablecappad

  \begin{center}
  {\tablesize

\newcommand{\mymidrule}{\midrule}

\begin{tabular}{
    S[table-format=2,table-space-text-post={\tabm{}}]
    cc
    S[table-format=1,table-space-text-post={\tabA{}}]
    S[table-format=2,table-space-text-post={\tabuA{}}]
    S[table-format=2,table-space-text-post={\tabmW{}}]
    S[table-format=4,table-space-text-post={\tabus{}}]
    S[table-format=4,table-space-text-post={\tabus{}}]
    S[table-format=4,table-space-text-post={\tabus{}}]
    S[table-format=3.1,table-space-text-post={\tabkbps{}}]
  }
  \toprule
  {\textbf{Distance}}       &
  \multicolumn{2}{c}%
  {\textbf{Target Circuit}} &
  {\textbf{Laser}}          &
  \multicolumn{2}{c}%
  {\textbf{At the Target}}  &
  \multicolumn{3}{c}%
  {\textbf{Configuration}}  &
  {\textbf{Data rate}}     \\
  \cmidrule(rl){2-3} \cmidrule(rl){4-4} \cmidrule(rl){5-6} \cmidrule(rl){7-9}
                            &
    {Resistor}              &
    {Capacitor}             &
    {Input Current}         &
    {Optical Power}         &
    {Current}               &
    {$\bm{\tone}$}          &
    {$\bm{\tzero}$}         &
    {$\bm{\toff}$}          &
    {}                     \\
  \mymidrule
     \tabm{10} & \CIRCLE && \tabA{1} & \tabmW{12} & \tabuA{37} & \tabus{  40} & \tabus{  15} & \tabus{  15} & \tabkbps{18.2} \\ %
     \tabm{20} & \CIRCLE && \tabA{2} & \tabmW{58} & \tabuA{43} & \tabus{  40} & \tabus{  15} & \tabus{  15} & \tabkbps{18.2} \\
     \tabm{25} & \CIRCLE && \tabA{2} & \tabmW{37} & \tabuA{20} & \tabus{  40} & \tabus{  15} & \tabus{  15} & \tabkbps{18.2} \\
     \tabm{30} & \CIRCLE && \tabA{4} & \tabmW{50} & \tabuA{32} & \tabus{  40} & \tabus{  15} & \tabus{  15} & \tabkbps{18.2} \\
     \tabm{35} & \CIRCLE && \tabA{4} & \tabmW{45} & \tabuA{35} & \tabus{  50} & \tabus{  15} & \tabus{  25} & \tabkbps{13.3} \\
     \tabm{40} & \CIRCLE && \tabA{4} & \tabmW{35} & \tabuA{20} & {    --    } & {    --    } & {    --    }&     { \xmark }  \\
  \mymidrule
     \tabm{35} && \CIRCLE & \tabA{4} & \tabmW{45} & \tabuA{35} & \tabus{3800} & \tabus{2100} & \tabus{1200} & \tabbps{200}   \\
     \tabm{40} && \CIRCLE & \tabA{4} & \tabmW{35} & \tabuA{20} & \tabus{3800} & \tabus{2100} & \tabus{1200} & \tabbps{200}   \\
  \bottomrule
\end{tabular}

}
  \end{center}
  \tablepad%
\end{table*}

\begin{table*}[b]
  \caption{Achievable data rates of different target devices
  (WLAN router, telephone, micro computer) at \SI{30}{\cm} distance.}
  \label{tab:pre_exp_rates}
  \tablecappad

  \begin{center}
  {\tablesize

\newcommand{\mymidrule}{\cmidrule(rl){1-3}\cmidrule(rl){4-7}\cmidrule(rl){8-11}}

\begin{altfootnotes}
\begin{tabular}{
    l
    l<{\hspace{-3mm}}l
    ll
    r<{\hspace{-3mm}}l
    S[table-format=3,table-space-text-post={\tabus{}}]
    S[table-format=3,table-space-text-post={\tabus{}}]
    S[table-format=3,table-space-text-post={\tabus{}}]
    S[table-format=5,table-space-text-post={\tabbps{}}]
  }
  \toprule
    {\textbf{Target device}} &
    \multicolumn{2}{l}
    {\textbf{Processor}}     &
    {\textbf{Laser}}         &
    {\textbf{LED}}           &
    \multicolumn{2}{l}
    {\textbf{GPIO}}          &
    {$\bm{\tone}$}           &
    {$\bm{\tzero}$}          &
    {$\bm{\toff}$}           &
    {\textbf{Data rate}}    \\
  \mymidrule
    \tlmr                  & Atheros AR-9331 &  (\SI{400}{\mega\hertz}) & green  & green        &   0 & (WiFi LED)         & \tabus{200} & \tabus{100} & \tabus{100} & \tabbps{ 3333} \\
    \tlwr                  & Atheros AR-9132 &  (\SI{400}{\mega\hertz}) & violet & green        &   5 & (QSS LED)          & \tabus{150} & \tabus{ 75} & \tabus{100} & \tabbps{ 4000} \\
    \rpi                   & BCM2837B0       & (~\SI{1.4}{\giga\hertz}) & violet & green\footnote[2]{}  
                                                                                                &  26 & (Pin Header)       & \tabus{ 30} & \tabus{ 15} & \tabus{ 15} & \tabbps{22222} \\
  \mymidrule
    \yealink               & DSPG DVF-9918   &  (\SI{400}{\mega\hertz}) & violet & green        & 112 & (green/red button) & \tabus{700} & \tabus{350} & \tabus{300} & \tabbps{ 1000} \\
    \rpi                   & BCM2837B0       & (~\SI{1.4}{\giga\hertz}) & violet & green\footnote[2]{}
                                                                                                &  26 & (Pin Header)       & \tabus{320} & \tabus{180} & \tabus{180} & \tabbps{ 2000} \\
    (with \SI{10}{\nano\farad} capacitor) &&&&&&&\\
  \bottomrule\\
    \multicolumn{5}{l}
    {\footnote[2]{} Using the LEDs of the \yealink telephone.}
\end{tabular}
\end{altfootnotes}

}
  \end{center}
  \tablepad%
\end{table*}

\subsection{Data rate}
\label{sec:sending-results}

To assess the achievable data rate of our attack, we measure
transmission in two different settings: We first experiment in a
laboratory setting to optimize our setup over a short distance and
second, in a realistic setting across long distances to demonstrate the
feasibility of the attack in practice.

\subsubsection{Short-distance transmission}
\label{sec:short}

In the first experiment, we measure the transmission per target device
over a distance of \SI{30}{\cm}. This is too short for a practical
instantiation of the attack, but enables us to optimize our setup for
subsequent long-distance transmissions.
Moreover, by narrowing down the external influence, we are able to
establish an empirical upper limit that may be achieved with the
particular hardware of the attacked devices. The used components (LEDs,
GPIOs specifications, processor, etc.) and the achievable data rate are
summarized in \autoref{tab:pre_exp_rates}. The specified times indicate
the configuration for the pulse-width modulation~(PWM).
In addition to the three target devices, we also include a \rpi as a
reference device.

A few things stand out: First, for the \tlmr, we make use of a green
laser rather than the more powerful engraving laser. This was necessary,
as the device's LED emits green light, in contrast to the others that
emit white light but carry a green cap. The green laser, however, is
slower than the purple one such that we yield a lower data rate.
Second, both devices are \num{5.5}$\times$ slower than the \rpi, which
underlines a certain dependency of the data rate on the CPU speed and,
thus, the possible sampling rates.
Third, the \Yealink is significantly slower than all the above. As
mentioned earlier, for this device an additional capacitor is built-in
such that the sampling rate is limited by the charge time of the
capacitor. Equipping the \rpi with the very same capacitor and the same
LEDs enables twice the data rate, due to the higher sampling rate.

To expand on these results and stretch the limits of the attack,
subsequently, we focus on the LEDs of the \Yealink in combination with
the \rpi that is equipped with a \SI{1.4}{\giga\hertz}~CPU.

\subsubsection{Long-distance transmission}
\label{sec:long}

For our long-distance experiment, we consider both predominate circuit
types that either use a series-connected resistor, as used by the \tlwr
router, or an additional capacitor, as it is the case for the
\yealink telephone. The setup remains as depicted in
\autoref{fig:setup_attack}.

We transmit data packets of \SI{1000}{\byte} in size and record them on
the other end. The recorded data is then compared with the transmitted
one to verify error-free communication.
These transmissions are evaluated for different circuits and with
increasing distances for \num{10}, \num{20}, \num{30}, \num{35}, and
\SI{40}{\meter}. \autoref{tab:recv} summarizes the results.
\emph{All measurements have been conducted indoors, due to safety reasons.}

Over \num{10}~consecutive repetitions of each experiment, data
transmission succeeds without a \mbox{single bit error ($\mathit{BER}$ $
= \perc{0}$)}.
For circuits with a series-connected resistor (\tlwr and \tlmr routers),
we achieve a remarkable data rate of \SI{18.2}{\kbps} across \SI{30}{m}.
Beyond this, the rate declines to \SI{13.3}{\kbps} for \SI{35}{m} and
transmission comes to a halt at \SI{40}{m}.
Similarly to the short-distance measurements before, we however also see
a significant difference in the achievable data rate for targets that
use a capacitor, which are slower by a factor of \num{100}.
Nevertheless, even \SI{200}{\bps} are sufficient to establish a
command-and-control channel to air-gapped systems and enable
orchestrating different malicious activities.
Moreover, the capacitor enable to operate on lower levels of induced
current at the target and, thus, allows to overcome larger distances.

To better highlight these relations, we also measure the optical power
that reaches the target with a Thorlabs~PME320E optical power meter and
the corresponding Thorlabs~S120VC power sensor.
To match the size of an SMD LED and its effective area, we cover large
portions of the sensor such that only about \SI{1}{\milli\meter\squared}
remains sensitive.
Even with a laser input current of \SI{4}{\ampere} the induced current
only reaches \SI{20}{\micro\ampere} in \SI{40}{\meter} distance, while
for shorter ranges at least \SI{30}{\micro\ampere} are reached.
In this setting, \SI{20}{\micro\ampere} is a tipping point, where a
reliable, error-free communication channel can still be established  for
circuits with series-connected resistors. For large distances, however,
vibrations are the limiting factor that impair the attack. This may be
counteracted using fast steering mirrors to level the laser beam as
described in \autoref{sec:vibrations}.
Targets that incorporate circuits with capacitors can even be
communicated with across \SI{40}{\meter}.
This is equivalent to the width of a highway with eight lanes including
median and shoulders on each side.
Charging the capacitor is largely unaffected by vibration, but slightly
slowed down. This, however, is easily compensated by the communication
protocol~(\cf~\autoref{sec:protocol}).

\ourresult{Summary} We demonstrate data transmission to built-in LEDs
with data rates of \SI{18.2}{\kbps} and up to \SI{30}{\meter} if no
capacitor is used in the circuit. Consequently, infiltrating data into
air-gapped devices, such as the two TP-Link routers, becomes possible at
speeds comparable to regular~modems.%

\section{Exfiltrating Data}%
\label{sec:attack-receive}

We continue to show how data can be exfiltrated from the targeted
devices. By flashing built-in LEDs, it is possible to establish
arbitrary communication protocols and transmit data.
Reaching realistic attack distances, however, requires efficient optical
equipment. In contrast to prior work, %
we show that it is perfectly feasible to overcome large
distances~\emph{and}~achieve high data rates %
without modifications of the targeted hardware.

We begin with a systematic evaluation of the sending capabilities of
LEDs in~\autoref{sec:leds-send}. Subsequently,
in~\autoref{sec:exfiltrate-cam,sec:exfiltrate-apd}, we demonstrate data
exfiltration in two different scenarios, where we use a)~consumer
high-speed cameras as available in modern smartphones, and b)~more
advanced, but equally affordable avalanche photodetectors.
With the latter it is possible to push the limits of the
communication and significantly outperform existing covert
channels that use~LEDs.

\begin{figure*}[t]\vspace*{-2mm}
  \newcommand{\myoffset}{\hspace*{-10mm}}
  \centerline{
  \subfloat[\myoffset]
  {\includegraphics[width=0.33\textwidth]{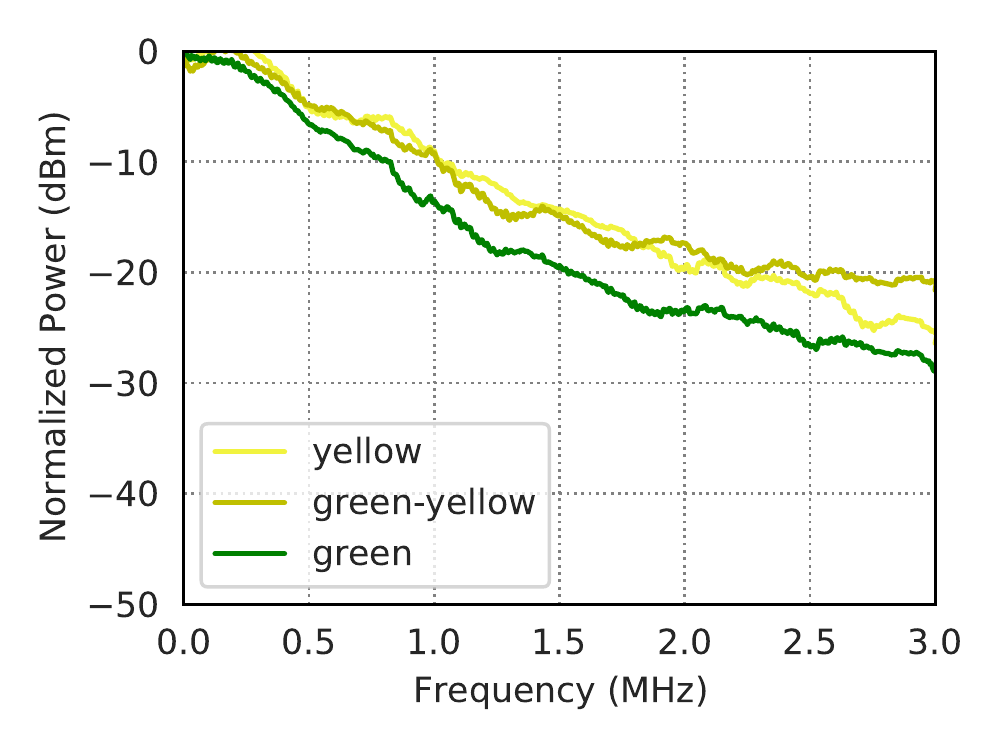}}
  \subfloat[\myoffset]
  {\includegraphics[width=0.33\textwidth]{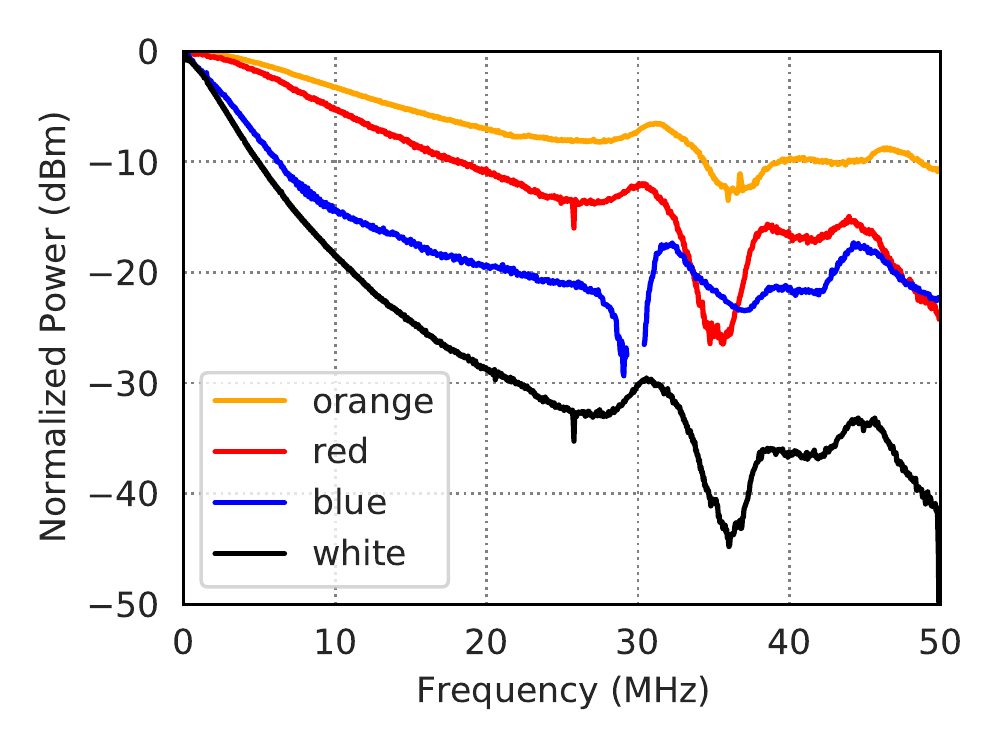}}
  \subfloat[\myoffset]
  {\includegraphics[width=0.33\textwidth]{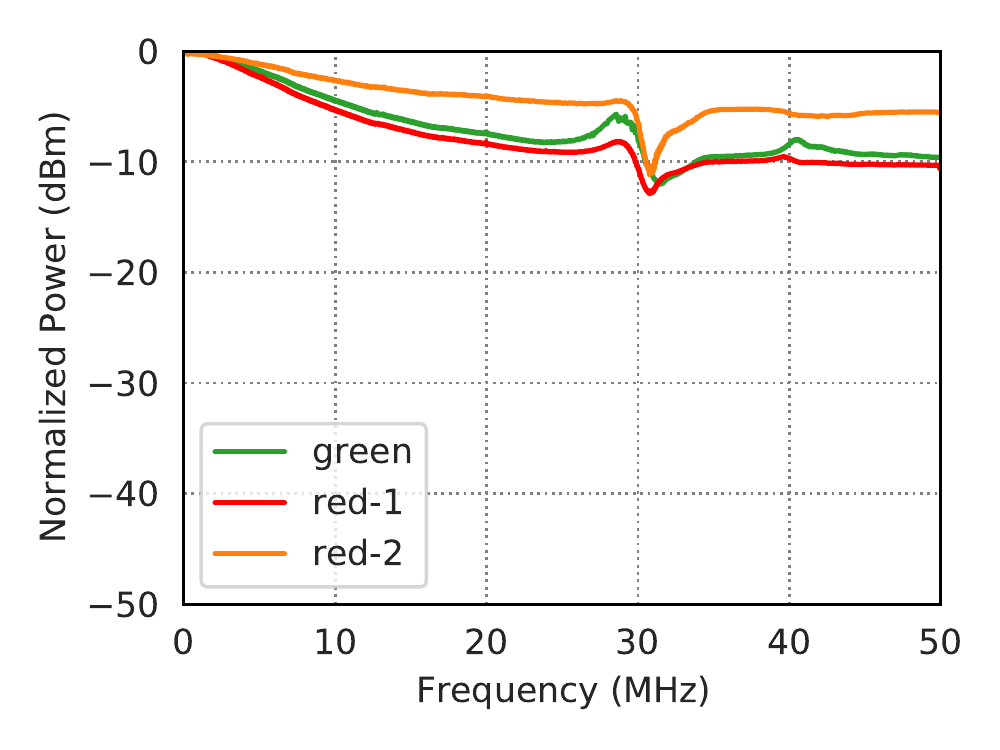}}
  }
  \figurecappad\vspace*{-2mm}
  \caption{Frequency response of a)~low-bandwidth and
  b)~high-bandwidth conventional
  {SMD\protect\textsuperscript{\ref{fnote:smd}}} diodes, and c)~of those
  used in the \mbox{\yealink}. The response of the photodetector itself
  has been subtracted.}
  \label{fig:LED_BW}\vspace*{-1mm}
\end{figure*}

\subsection{LEDs as Sender}
\label{sec:leds-send}

For data exfiltration the emitted light has to be powerful enough to be
observable by the adversary.
Subsequently, we thus determine the characteristics of different LEDs
and measure their bandwidth, that is, the theoretically achievable data
rate, as well as the optical power of the emitted light
in~\autoref{sec:bandwidth,sec:optical-power}, respectively.
Details and specifications of the LEDs are provided in
\autoref{app:leds}.

\subsubsection{Bandwidth}
\label{sec:bandwidth}

For evaluating the theoretically achievable data rates that can be
transmitted with consumer LEDs, we proceed to measure the modulation
bandwidth. The setup is as follows:
The LEDs are actuated by a radio frequency generator~(RFG) that
generates a sweep frequency from \SI{100}{Hz} up to \SI{50}{MHz} with an
effective output voltage of \SI{3.4}{V}. The light emitted by the LED is
then measured with a photodetector~(PD), a Thorlabs PDA8A, in a
distance of \SI{0.5}{\cm}.  This silicon-based detector exhibits a
spectral response of \num{320}--\SI{1000}{\nm} on an active area of
\SI{0.8}{\mm\squared}. The wavelengths of the previously
characterized LEDs thus are entirely in range.
The actual response of the diodes is finally recorded with an electrical
spectrum analyzer~(ESA) in maximum hold~mode.

\begin{table}[!b]
  \tablepad\vspace*{-1mm}
  \caption{Modulation bandwidth and optical output power for different consumer LEDs.}
  \label{tab:LED-power}
  \tablecappad

  \begin{center}
  {\tablesize

\newcommand{\mymidrule}{\cmidrule(rl){1-2}\cmidrule(rl){3-6}}

\begin{tabular}{
    cl
    S[table-format=2.2,table-space-text-post={\tabMHz{}}]
    S[table-format=4,table-space-text-post={\tabuW{}}]
    S[table-format=4,table-space-text-post={\tabuW{}}]
    S[table-format=4,table-space-text-post={\tabuW{}}]
  }
  \toprule
  \multicolumn{2}{c}
  {\textbf{LED}} &
  {\textbf{Bandwidth}} &
  \multicolumn{3}{c}%
  {\textbf{Optical Power}} \\
  \cmidrule(rl){4-6}
    & & {[3\,dB]} & { Bias } & { Sine } & { Burst }\\
  \mymidrule
    \parbox[t]{2mm}{\multirow{7}{*}{\rotatebox[origin=c]{90}{Exemplary selection}}}
    & yellow & \tabMHz{ 0.42} & \tabuW{  44} & \tabuW{  29} & \tabuW{  17} \\
    & gr/ye  & \tabMHz{ 0.41} & \tabuW{  12} & \tabuW{   7} & \tabuW{   5} \\
    & green  & \tabMHz{ 0.33} & \tabuW{  55} & \tabuW{  29} & \tabuW{  16} \\
    & orange & \tabMHz{ 9.55} & \tabuW{ 629} & \tabuW{ 786} & \tabuW{ 186} \\
    & red    & \tabMHz{ 6.73} & \tabuW{1690} & \tabuW{1290} & \tabuW{ 212} \\
    & blue   & \tabMHz{ 2.06} & \tabuW{5580} & \tabuW{2780} & \tabuW{1470} \\
    & white  & \tabMHz{ 1.75} & \tabuW{5800} & \tabuW{3930} & \tabuW{1580} \\
  \mymidrule
    \parbox[t]{2mm}{\multirow{3}{*}{\rotatebox[origin=c]{90}{Yealink}}}
    & green  & \tabMHz{ 7.22} & \tabuW{ 258} & \tabuW{ 311} & \tabuW{ 119} \\
    & red-1  & \tabMHz{ 6.04} & \tabuW{ 551} & \tabuW{ 640} & \tabuW{ 152} \\
    & red-2  & \tabMHz{11.22} & \tabuW{ 257} & \tabuW{ 646} & \tabuW{ 131} \\
  \bottomrule
\end{tabular}

}
  \end{center}
  \tablepad
\end{table}

\autoref{fig:LED_BW} shows the frequency response for a)~low-bandwidth
and b)~high-bandwidth conventional SMD LEDs, and c)~for those used by the
\Yealink.
The conventional SMD diodes show widely different behavior concerning
the possible modulation bandwidth, which is founded in the used
materials and construction scheme of the devices.
Usually, AlGaInP and InGaN are used in commercially available LEDs for
red, amber and yellow as well as green and blue colors,
respectively~\citep{Kasap2013, Schubert2018}. Moreover, InGaN diodes
typically are constructed as \emph{Multiple Quantum Well}~(MQW)
structures, whereas AlGaInP devices are build as \emph{Double
  Heterojunctions}~(DH).
Due to the different materials and layer structures, there are diverse
charge carrier lifetimes, which in turn results in the different
modulation bandwidths~\citep{pei2013led}.

The maximum theoretical modulation bandwidth of LEDs is limited to
\SI{2}{GHz}~\citep{chen1999ghz, Koudelka2011}.
In our experiments, the red and orange LEDs can reach data rates of
\num{7}~to~\SI{9}{MHz} for on-off modulation. The LEDs from the
Yealink telephone show even higher modulation bandwidths with rates of
up to \SI{11.2}{MHz}.
An overview of the exact bandwidth characteristics is given in
\autoref{tab:LED-power}. Values are specified as the \emph{full width
  at half maximum}~(FWHM) bandwidth, that is, the frequency where the
transmitted power has decreased by the half or \SI{3}{dB}.

\subsubsection{Optical power}
\label{sec:optical-power}

Finally, we measure the maximally emitted optical power for each LED to
assess how well consumer diodes are visible in distance. To this
end, we replace the photodetector %
with an optical power meter (Thorlabs~PME320E) and a power sensor
(Thorlabs~S120VC),
and measure the emitted power in three different settings: First, we
determine the \emph{maximum bias voltage} that may be applied. Second,
we modulate a \emph{sine wave} with a frequency of \SI{30}{\kilo\hertz}
with a voltage of \SI{5}{V}. Finally, we apply a burst of
\SI{16}{\bit} large data blocks at a rate of \SI{30}{\kbps} and repeat
it with \SI{500}{\hertz}, which allows us to estimate the power for
data transmission.
The results are shown in \autoref{tab:LED-power}.

In particular, the blue and white LEDs show the highest output powers
under the given conditions. However, the average output power is
decreased if modulation is applied, as we turn off the LED for
transmitting a digital zero. This, of course, needs to be considered for
determining the maximum distance and data rate in subsequent
experiemnts.
Generally speaking, the higher the output power the better we are able
to bridge large distances. Much of the restrictions imposed at this
point, however, can be overcome by hardware for effectively capturing
light as detailed in \autoref{sec:exfiltrate-apd}.

\subsection{Exfiltration using high-speed cameras}
\label{sec:exfiltrate-cam}

Modern smartphones often have the ability to capture slow-motion videos.
This boils down to an increased frame rate of the recorded video, that
is, high-speed camera functionality. These cameras are very sensitive to
small and dark light sources, such that they are perfectly suited for
our attack.

In the first experiment, we thus use an iPhone~11 that comes with a
\SI{240}{\fps} 1080p camera to exfiltrate data from three target
devices:
\tlwr, \tlmr, and the \yealink telephone. The procedure is as easy as
capturing a video of the target device while it is transmitting.
Analyzing the video stream, in turn, is done offline on more performant
hardware. In a more specialized setting, as for instance demonstrated in
\autoref{sec:exfiltrate-apd}, this can however be equally conducted
online.
Moreover, here the attacker is not limited to a specific sampling
strategy, such that we can use on-off-keying~(OOK) for modulation
(\cf~\autoref{sec:protocol}).
Due to the camera's limited frame rate and the Nyquist–Shannon sampling
theorem~\citep[see][]{KenDav92} the achievable data rate is limited to
\SI{120}{\bps}. To empirically verify this, we transmit three randomly
generated data chunks of \SI{500}{\byte} in size using each target
device, while filming it with the iPhone. The size of transmitted data
is reduced in comparison to the previous experiment as the iPhone's
internal storage quickly runs full given the large size of the
recorded~video.

\begin{table}[htbp]
  \tablepad
  \caption{Data rates using a \SI{240}{\fps} high-speed camera as
    receiver in two settings: \SI{2}{\meter} indoors and \SI{40}{\meter}
    outdoors.}
  \label{tab:camera_rates}
  \tablecappad

  \begin{center}
  {\tablesize

\begin{tabular}{
    l
    c
    S[table-format=3.2,table-space-text-post={\tabbps{}}]
  }
  \toprule
  { \textbf{Target device} } &
  { \textbf{Distance} } &
  { \textbf{Data rate} }\\
  \midrule
    \tlmr    & \tabm{2 -- 40} & \tabbps{119.05} \\
    \tlwr    & \tabm{2 -- 40} & \tabbps{119.05} \\
    \yealink & \tabm{2 -- 40} & \tabbps{119.05} \\
  \bottomrule
\end{tabular}

}
  \end{center}
\end{table}

We conduct two different sets of experiments, that are summarized in
\autoref{tab:camera_rates}. At first, we measure transmission on a short
range of \SI{2}{\meter} indoors, and proceed to long-range measurements
across \SI{40}{\meter} outdoors (the maximum distance possible at our
testing grounds). In line with the ``consumer setting'' of using
non-specialized hardware, for the second experiment, we use regular
binoculars (Minox \mbox{BL 10x44 HD}) to zoom in on the LEDs of the
target devices. In all cases, we nearly yield the theoretical maximum.
In contrast to transmitting data using laser beams, here a distance of
\SI{40}{\meter} does not affect the data rate. Larger distances could
not be investigated due to the boundaries of our testing grounds.
However, reception is possible as long as at least a single pixel that
represents the LED~is~visible.

\subsection{Exfiltration using photodetectors} %
\label{sec:exfiltrate-apd}

While high-speed cameras are very sensitive to light, they have a clear
limit imposed by their frame rate. Photodetectors, in turn, allow to
improve upon this limitation at the expense of sensitivity.
To yield the highest possible signal-to-noise ratio at rather low
optical input power levels, we use an avalanche photodetector~(APD) that
we characterized in \autoref{sec:photodiodes}. To compensate for large
distances and improve reception, an attacker may use more efficient
optical equipment.
In subsequent experiments, we thus employ the telescope, that has also
been used for focusing the laser beam on the target (the Navitar
Zoom~7000) rather than ordinary binoculars.
In \autoref{sec:exfiltrate-apd-datarate}, we again direct our attention
to the \Yealink with its green and red SMD~LEDs to inspect
a)~the raw observability in distance, and b)~the data rate in a
realistic~setting.

\begin{figure}[b]
  \vspace*{-4mm}
  \centerline{
    \includegraphics[width=0.9\columnwidth, height=110pt]{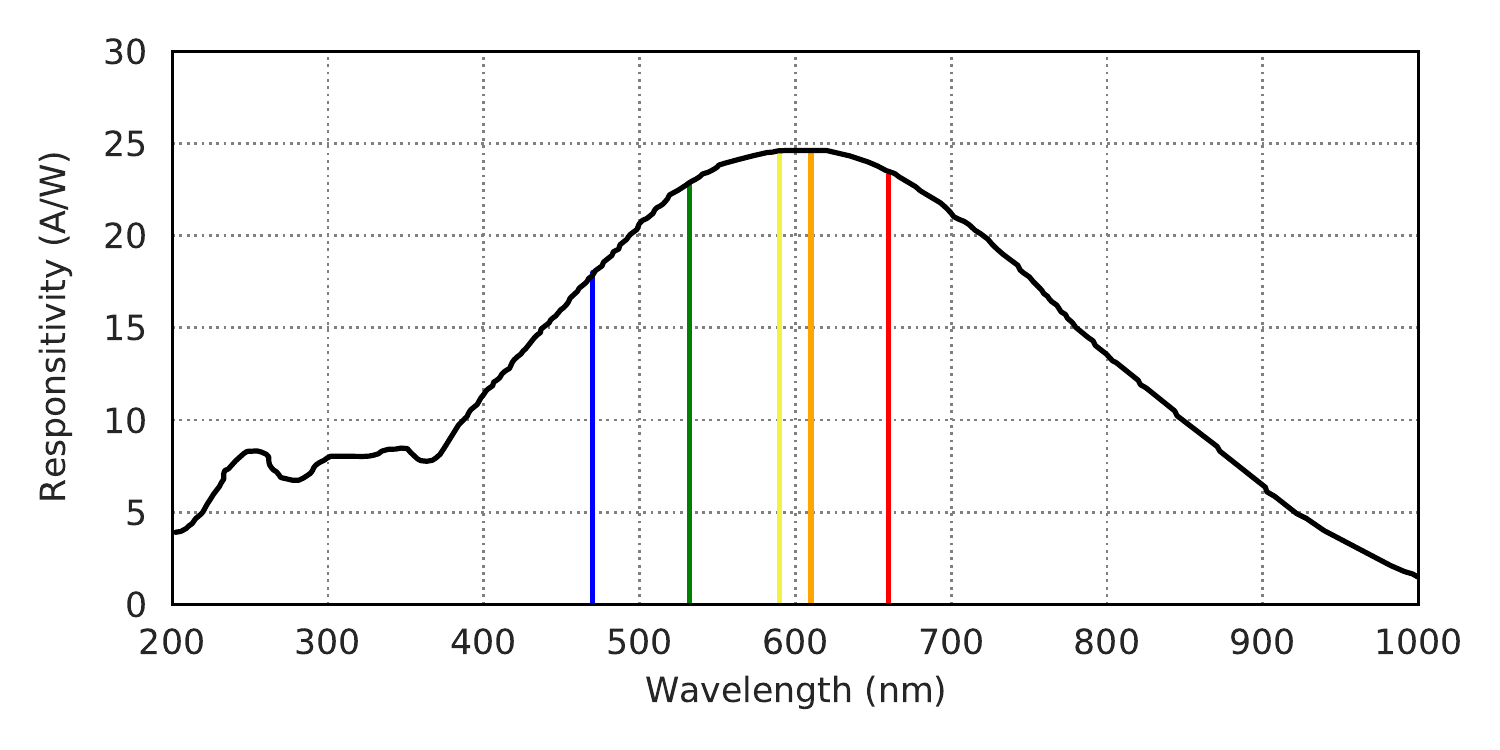}
  }
  \figurecappad\vspace*{-3mm}
  \caption{Responsivity of the APD440A2 photodetector at a gain factor of $M=50$.}
  \label{fig:APD440}
  \label{fig:APD-Character}
  \vspace*{-1mm}
\end{figure}

\subsubsection{Sensitivity of Photodetectors}
\label{sec:photodiodes}

For recording the data send out by an LED of a compromised device over
large distances, we require
highly sensitive and fast hardware to capture light.
Photodetectors are made from element semiconductors such as silicon,
germanium, or compound semiconductors such as indium gallium arsenide.
For visible light (\num{380}{nm}--\SI{780}{nm}) mainly detectors made of
silicon (\num{190}--\SI{1100}{nm}) and germanium
(\num{400}--\SI{1700}{nm}) are used. Due to the larger band gap of
silicon it is possible to also achieve comparable low noise.
Conventional photodetectors, however, have limited sensitivity and no
internal gain, such that additional transimpedance amplifiers are
necessary for operation, which again reduces the overall \mbox{signal-to-noise}
ratio.
For measuring the smallest amounts of light so-called avalanche
photodetectors~(APDs) may be used, which produce a gain factor in the
hundreds using a photoelectric effect based on \mbox{impact ionization}.

To characterize the receiving capabilities of an attacker, we hence
evaluate the sensitivity of a silicon-based APD. In particular, we make
use of the Thorlabs APD440A2, which promises a low signal-to-noise ratio
at rather low optical input power levels. It operates on a range from
\num{200}--\SI{1000}{\nm} with a maximum responsivity of
\SI{25}{\ampere\per\watt} at a noise-equivalent power of
\SI{2.5}{\femto\watt}.
As we attempt to measure a wide-range of light-emitting diodes with
different wavelengths, we break down the responsivity of the APD by
color in \autoref{fig:APD440}. Especially the green and red LEDs exhibit
almost optimal output levels, making the avalanche photodetector a
well-suited receiver for exfiltrating data from consumer devices.
For subsequent experiments, we use the photodetector with a gain or
multiplication factor of $M=50$. More information on this configuration
and the exact frequency response is provided in \autoref{app:gain}.

\subsubsection{Data rate}
\label{sec:exfiltrate-apd-datarate}

Finally, we measure the data rate when exfiltrating data using LEDs. For
this, we implement the experimental setup depicted
in~\autoref{fig:setup_receive} and conduct two different measurements.
First, we gauge the electrical response of the photodetector to LEDs in
distance to explore the limits of the communication channel.
For this, we modulate the LEDs using an arbitrary waveform
generator~(AWG) such that the diodes emit a burst of rectangular signals.
Second, we replace the AWG with the malicious implant at the target
and transmit large data chunks to determine the bit error rate of
the communication~channel.

\begin{figure}[htbp]
\centerline{\includegraphics[width=\columnwidth]{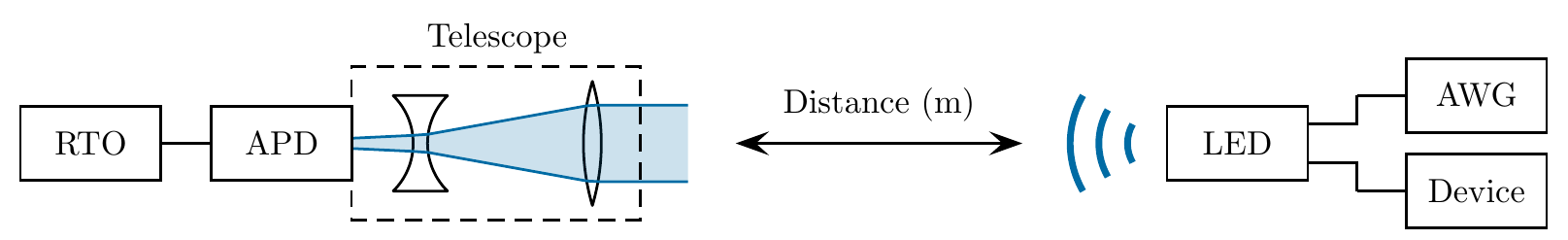}}

\figurecappad
\caption{Experimental setup for exfiltrating data in two
different setting: First, driven by an arbitrary waveform
generator~(AWG) and second, operated by the targeted device. Signals are
captured with an avalanche photodetector~(APD) and processed with a
real-time oscilloscope~(RTO).}
\label{fig:setup_receive}
\end{figure}

\begin{table}[b]
  \caption{Voltage levels of data streams captured using an avalanche photodetector a)~with and b)~without resistor.}
  \label{tab:voltage}
  \tablecappad

  \begin{center}
  \subfloat[with resistor]{\tablesize

\newcommand{\mymidrule}{\cmidrule(rl){1-1} \cmidrule(rl){2-7}}

\begin{tabular}{
    l
    S[table-format=1.2,table-space-text-post={\tabV{}}]
    S[table-format=1.2,table-space-text-post={\tabV{}}]
    S[table-format=1.2,table-space-text-post={\tabV{}}]
    S[table-format=1.2,table-space-text-post={\tabV{}}]
    S[table-format=1.2,table-space-text-post={\tabV{}}]
    S[table-format=1.2,table-space-text-post={\tabV{}}]
  }
  \toprule
  {\textbf{LED}} &
  {\textbf{\SI{ 5}{\meter}}} &
  {\textbf{\SI{10}{\meter}}} &
  {\textbf{\SI{15}{\meter}}} &
  {\textbf{\SI{20}{\meter}}} &
  {\textbf{\SI{25}{\meter}}} &
  {\textbf{\SI{30}{\meter}}} \\
  \mymidrule
    green     & \tabV{0.23} & \tabV{0.11} & \tabV{0.06} & \tabV{0.03} & \tabV{0.02} & {\makebox[0pt][l]{~~~~\,\xmark}\phantom{\tabV{0.000}}} \\
    red-1     & \tabV{0.11} & \tabV{0.07} & \tabV{0.03} & \tabV{0.01} & { \xmark  } & {  \xmark  } \\
    red-2     & \tabV{0.06} & \tabV{0.02} & { \xmark  } & { \xmark  } & { \xmark  } & {  \xmark  } \\
  \bottomrule
\end{tabular}

}\\
  \subfloat[without resistor]{\tablesize

\newcommand{\mymidrule}{\cmidrule(rl){1-1} \cmidrule(rl){2-7}}

\begin{tabular}{
    l
    S[table-format=1.2,table-space-text-post={\tabV{}}]
    S[table-format=1.2,table-space-text-post={\tabV{}}]
    S[table-format=1.2,table-space-text-post={\tabV{}}]
    S[table-format=1.2,table-space-text-post={\tabV{}}]
    S[table-format=1.2,table-space-text-post={\tabV{}}]
    S[table-format=1.2,table-space-text-post={\tabV{}}]
  }
  \toprule
  {\textbf{LED}} &
  {\textbf{\SI{ 5}{\meter}}} &
  {\textbf{\SI{10}{\meter}}} &
  {\textbf{\SI{15}{\meter}}} &
  {\textbf{\SI{20}{\meter}}} &
  {\textbf{\SI{25}{\meter}}} &
  {\textbf{\SI{30}{\meter}}} \\
  \mymidrule
    green     & \tabV{2.26} & \tabV{0.83} & \tabV{0.41} & \tabV{0.20} & \tabV{0.14} & \tabV{0.10} \\
    red-1     & \tabV{0.89} & \tabV{0.30} & \tabV{0.13} & \tabV{0.09} & \tabV{0.07} & \tabV{0.02} \\
    red-2     & \tabV{0.13} & \tabV{0.03} & { \xmark  } & { \xmark  } & { \xmark  } & { \xmark  }\\
  \bottomrule
\end{tabular}

}
  \end{center}
  \tablepad
\end{table}

We begin by modulating bursts of \num{16} rectangular signals at a data
rate of \SI{30}{\kbps} and repeat this at \SI{500}{\hertz} with an
effective output voltage of \SI{3.3}{\volt}.
The signals captured by the APD are recorded with a real time
oscilloscope %
resulting in high/low voltage patterns that can be interpreted as the
corresponding data pattern. The measured voltage levels are reported in
\autoref{tab:voltage}.
We characterize the measured response a)~with and b)~without an
additional series-connected resistor of \SI{470}{\ohm} as used by the
\Yealink.
For both measurements, it is clearly visible that the observed levels
and thus the response decays with increasing distance. With the resistor
in place, we are able to bridge a distance of \SI{25}{\meter} with the
green LED and \SI{20}{\meter} with the first red LED~(\code{red-1}).
The second red diode~(\code{red-2}) is less bright and thus reaches
less far, which matches the intuition one has of the setting.
Subsequently, we thus use the telephone's green LED for our experiments
on the data rate in~practice.

We transmit \SI{1000}{\byte} large data chunks at an output voltage of
\SI{3.3}{\volt}, digitize the measured high/low levels, and calculate
the bit error rate~(BER) between sent and received bit streams. The
results are presented in~\autoref{tab:BER}.
For up to \SI{25}{\meter} and data rates of \SI{100}{\kbps},
transmission is possible with minimal bit errors ($\mathit{BER} =
\perc{0.1}$). Due to the limited bandwidth of the APD, however, the
measured bit error increases significantly for data rates of
\SI{200}{\kbps} and distances above \SI{25}{\meter}, such that
transmission abruptly becomes impossible.
With a more sensitive detector and improved optics, however, this can
likely be enhanced even further.

Moreover, one may even modulate the background light of the telephone's
display to transmit data. In comparison to the rather small LEDs, the
LCD display constitutes a large and bright light source that promises a
better reception at the photodetector. We extend on this idea in
\autoref{app:lcd}.

\begin{table}[htbp]
  \tablepad
  \caption{Bit Error Rates (BER) of transmissions from the target to the
  attacker at different data rates using the green LED of the \Yealink.}
  \label{tab:BER}
  \tablecappad

  \begin{center}
  {\tablesize

\begin{tabular}{
    S[table-format=2,table-space-text-post={\tabm{}}]
    S[table-format=1.1,table-space-text-post={\perc{}}]
    S[table-format=1.1,table-space-text-post={\perc{}}]
    S[table-format=1.1,table-space-text-post={\perc{}}]
    S[table-format=1.1,table-space-text-post={\perc{}}]
    S[table-format=2.2,table-space-text-post={\perc{}}]
  }
  \toprule
  {\textbf{Distance}}   &
  \multicolumn{4}{c}%
  {\textbf{Data rate}} \\
  \cmidrule(rl){2-5}
  {               }     &
  {\SI{  1}{\kbps}}     &
  {\SI{ 50}{\kbps}}     &
  {\SI{100}{\kbps}}     &
  {\SI{200}{\kbps}}    \\
  \midrule
     \tabm{ 5} & 0.0\perc{} & 0.0\perc{} & 0.0\perc{} &  0.1\perc{} \\
     \tabm{10} & 0.0\perc{} & 0.0\perc{} & 0.0\perc{} &  0.9\perc{} \\
     \tabm{15} & 0.0\perc{} & 0.0\perc{} & 0.0\perc{} &  2.2\perc{} \\
     \tabm{20} & 0.0\perc{} & 0.0\perc{} & 0.1\perc{} & {  \xmark } \\
     \tabm{25} & 0.0\perc{} & 0.0\perc{} & 0.1\perc{} & {  \xmark } \\
     \tabm{30} & { \xmark } & { \xmark } & { \xmark } & {  \xmark } \\
  \bottomrule
\end{tabular}

}
  \end{center}
  \tablepad\vspace*{-2mm}
\end{table}

\ourresult{Summary} We demonstrate that LEDs used in office devices can
be used for high-speed exfiltration of data. We achieve a throughput of
\SI{100}{\kbps} up to a distance of \SI{25}{\meter}. This data rate
enables to transfer megabytes of data within minutes and thus poses a
serious threat to air-gapped environments.%

\section{Conclusion}
\label{sec:conclusions}

An air-gapped system is unreachable from the outside by definition.
The research community, however, has shown that this is not
necessarily true, and has demonstrated multiple ways of bridging the
gap in the past. While these methods feature various very creative
covert channels, their practical utility often remains questionable
due to low data rates, short distances, or only unidirectional
communication.
We are the first to demonstrate the exfiltration of data at
\SI{100}{\kbps} over \SI{25}{\meter} \emph{and} allow for infiltrating
data to an unmodified device in \SI{30}{\meter} distance at
\SI{18.2}{\kbps}. With this, we show that covert channels are not bound
to obscure and rare settings, but are a real threat in practice.

Network operators of high-security facilities and industries that are
at risk of cooperate espionage must not settle for merely air-gapping
a system, but also need to prevent targeted attacks that make use of a
physical \mbox{component---for instance,} using optical channels that
require a direct line of sight. Obvious countermeasures for this
particular attack are optically opaque rooms. However, as demonstrated
by related approaches, this threat extends to other variations using
electromagnetic, acoustic, or power-dependent aspects.

\fi

\makeatletter
\if@ACM@nonacm\else
\begin{acks}
The authors gratefully acknowledge funding from the German Federal
Ministry of Education and Research~(BMBF) under the project
XXX~(FKZ~00XXX0000Y).
\end{acks}
\fi

\balance
{\small\footnotesize
  \bibliographystyle{abbrvnat} %
\iftrue

\fi
}
\makeatother

\clearpage

\begin{table*}[t]
  \tablepad
  \caption{Details and specification of the LEDs characterized in \autoref{sec:leds-recv,sec:leds-send}.}
  \label{tab:led_specs}
  \tablecappad

  \begin{center}
  {\tablesize

\newcommand{\mymidrule}{\cmidrule(rl){1-1}\cmidrule(rl){2-4}\cmidrule(rl){5-8}}

\begin{tabular}{
    llll
    S[table-format=3,table-space-text-post={\,mcd}]
    S[table-format=3,table-space-text-post={\si{\degree}}]
    S[table-format=2,table-space-text-post={\tabmA{}}]
    S[table-format=1.1,table-space-text-post={\tabV{}}]
  }
  \toprule
  {\textbf{Color}}              &
  {\textbf{Type}}               &
  {\textbf{Manufacturer}}       &
  {\textbf{Part No.}}           &
  {\textbf{Brightness}}         &
  {\textbf{Angle of Radiation}} &
  {\textbf{Current}}            &
  {\textbf{Voltage}}           \\
  \mymidrule
  Yellow       & SMD-LED 0603 & Würth Elektronik & 150060YS75000  & 120\,mcd & \tabdeg{140} & \tabmA{30} & \tabV{2.0} \\
  Green/Yellow & SMD-LED 0603 & Broadcom         & HSME-C191      &  50\,mcd & \tabdeg{170} & \tabmA{20} & \tabV{2.1} \\
  Green        & SMD-LED 0603 & Kingbright       & KPHCM-2012CGCK &  50\,mcd & \tabdeg{110} & \tabmA{20} & \tabV{2.1} \\
  Blue         & SMD-LED 0603 & TRU Components   & 1573646             & 120\,mcd & \tabdeg{120} & \tabmA{25} & \tabV{3.2} \\
  White        & SMD-LED 0603 & TRU Components   & 1573647  & 400\,mcd & \tabdeg{120} & \tabmA{25} & \tabV{3.2} \\
  Orange       & SMD-LED 0603 & Kingbright       & KP-1608SECK    & 180\,mcd & \tabdeg{120} & \tabmA{20} & \tabV{2.1} \\
  Red          & SMD-LED 0603 & Würth Elektronik & 150060RS75000  & 250\,mcd & \tabdeg{140} & \tabmA{30} & \tabV{2.0} \\
  \bottomrule
\end{tabular}

}
  \end{center}
  \tablepad
\end{table*}

\appendix
\section{Selection of LEDs}
\label{app:leds}

For evaluating the characteristics of light-emitting diodes and the
suitability of these to be used for a covert channel, we have used a
selection of SMD LEDs in \autoref{sec:leds-recv,sec:leds-send}. To allow for easy
reproducibility of our experiments, \autoref{tab:led_specs} list
manufacturers, part numbers, and basic properties of these.

\section{Photodetector Gain}
\label{app:gain}

\begin{figure}[htbp]
  \centerline{
    \includegraphics[width=0.9\columnwidth, height=110pt]{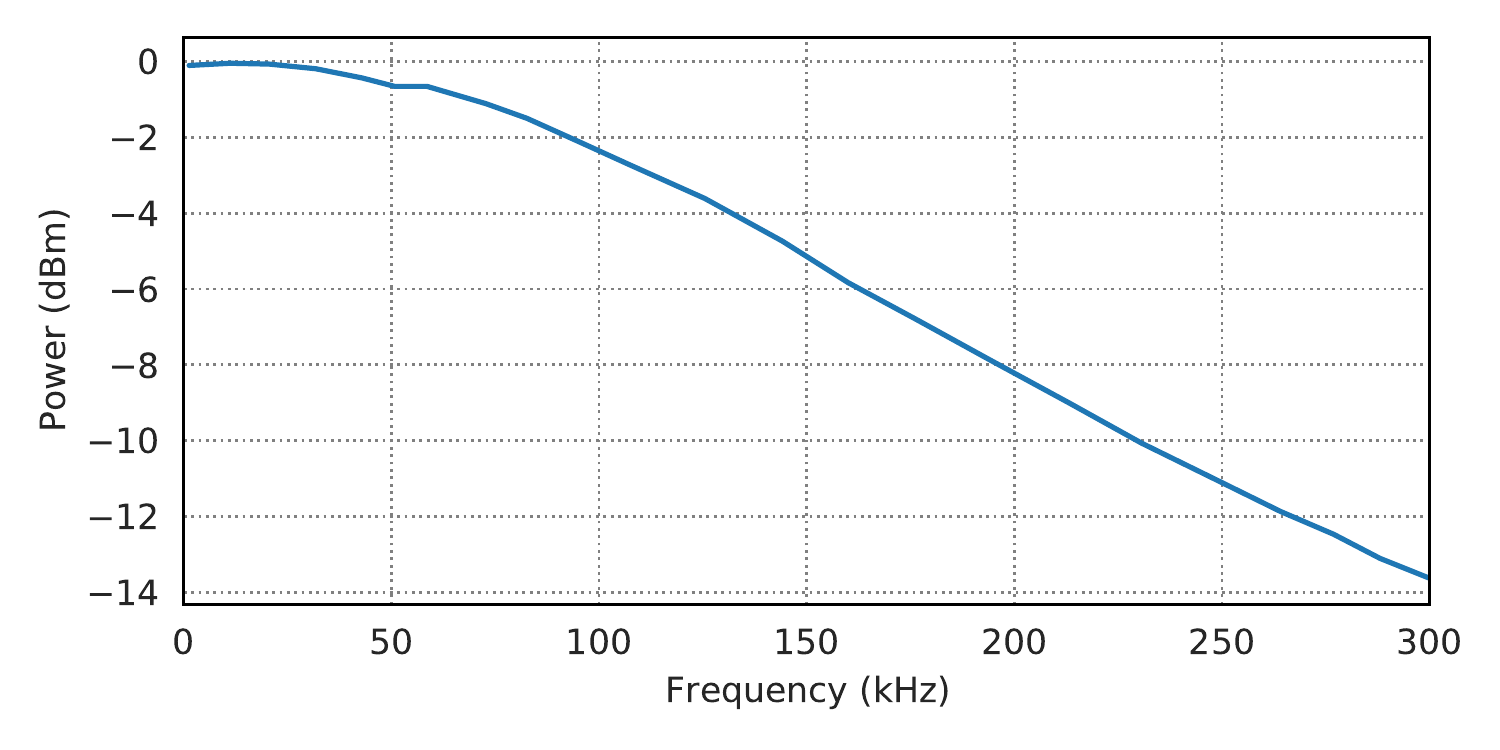}
  }

  \figurecappad\vspace*{-3mm}
  \caption{Frequency response of the APD440A2 photodetector with a gain factor $M=50$, operated at room temperature.}
  \label{fig:APD-freq}\vspace*{-3mm}
\end{figure}

A photodetector's gain or multiplication factor~$M$ is dependent on the
reverse bias voltage, that is used to create the electric field for
triggering the avalanche effect, but also the temperature. While the
multiplication factor increases and decreases with the reverse bias
voltage, it is inversely proportional to the temperature, meaning, the
gain increases at low temperatures, but decreases if the temperature
rises.
However, the amplification limits the FWHM bandwidth of the detector to
\SI{100}{\kHz}, which in turn bounds the maximally transmittable data
rate. For the experiments presented in \autoref{sec:exfiltrate-apd}, we
use a gain factor of $M=50$ which exhibits the frequency response shown
in~\autoref{fig:APD-freq}.

\begin{table}[b]
  \tablepad
  \caption{Bit Error Rates (BER) when exfiltrating data using the
  display of the \Yealink.}
  \label{tab:BER-display}
  \tablecappad

  \begin{center}
  {\tablesize

\begin{tabular}{
    S[table-format=2,table-space-text-post={\tabm{}}]
    S[table-format=1.1,table-space-text-post={\perc{}}]
    S[table-format=1.1,table-space-text-post={\perc{}}]
    S[table-format=1.1,table-space-text-post={\perc{}}]
    S[table-format=1.1,table-space-text-post={\perc{}}]
    S[table-format=2.2,table-space-text-post={\perc{}}]
  }
  \toprule
  {\textbf{Distance}}   &
  \multicolumn{4}{c}%
  {\textbf{Data rate}} \\
  \cmidrule(rl){2-5}
  {               }     &
  {\SI{  1}{\kbps}}     &
  {\SI{ 50}{\kbps}}     &
  {\SI{100}{\kbps}}     &
  {\SI{200}{\kbps}}    \\
  \midrule
     \tabm{ 5} &  0.0\perc{} & 0.0\perc{} & 0.0\perc{} & { \xmark } \\
     \tabm{10} &  0.0\perc{} & 0.0\perc{} & 0.0\perc{} & { \xmark } \\
     \tabm{15} &  0.0\perc{} & 0.0\perc{} & 0.0\perc{} & { \xmark } \\
     \tabm{20} & 26.0\perc{} & { \xmark } & { \xmark } & { \xmark } \\
     \tabm{25} & {  \xmark } & { \xmark } & { \xmark } & { \xmark } \\
  \bottomrule
\end{tabular}

}
  \end{center}
  \tablepad
\end{table}

\section{Transmitting Data with\\LCD Displays}
\label{app:lcd}

Similar to LEDs, the background light of an (LCD) display can be used to
transmit data. The display constitutes a large and rather bright light
source that promises a better reception at the APD.
However, next to brightness itself, also the wavelength of light is
crucial for the sensitivity of the (avalanche) photodetector. In case of
the \yealink, the background light mainly emits blue light, which is not
covered well by the detector as shown in \autoref{fig:APD440}.
As the APD responds less to blue than green light, this also reflects in
the distance that can be overcome.

Although the detector is not very well suited for the white/blue light
of the telephone's display, it still is possible to also exfiltrate data
using the background light. As shown in \autoref{tab:BER-display}, for a
distance of up to \SI{15}{\meter}, we achieve a data rate of
\SI{100}{\kbps} without any bit error when modulated with traditional
on-off keying.
Even at a distance of \SI{20}{\meter} transmission is possible with up
to \SI{1}{\kbps} and a BER of \perc{26}. Beyond this, however, no
noteworthy output and thus data rate can be~measured.

\end{document}